\documentclass[aps,prc,reprint,groupedaddress,nofootinbib]{revtex4-2}

\usepackage[utf8]{inputenc}
\usepackage{graphicx}
\usepackage{amsmath}
\usepackage{xcolor}
\usepackage{hyperref}
\hypersetup{hidelinks}

\begin{document}

\newcommand*{\Tmax}{T_{0,\textrm{max}}}
\newcommand*{\Teff}{T_{\rm eff}}
\newcommand{\sqrts}{\sqrt{s_{\textrm NN}}}
\newcommand{\tofix}[1]{\textcolor{red}{[To fix: #1]}}

\title{Electromagnetic estimate of the temperature of quark-gluon plasma produced in central ultrarelativistic nuclear collisions}
\author{Jean-Fran\c{c}ois Paquet}
\affiliation{Department of Physics and Astronomy, Vanderbilt University, Nashville, TN 37212, USA}
\author{Steffen A. Bass}
\affiliation{Department of Physics, Duke University, Durham, NC 27708-0305, USA}
\date{\today}

\begin{abstract}
Ultrarelativistic collisions of large nuclei produce a short-lived plasma of deconfined quarks and gluons. Of all the GeV-energy photons detected in these nuclear collisions, only a small number are emitted directly by the quark-gluon plasma.
The characteristic near-exponential energy spectrum of these photons is often associated with an effective temperature whose interpretation is complicated by spacetime averaging, Doppler shifts and other factors.
We show that the Doppler shift's effect on the slope of the photon spectrum is generally modest and can either increase or decrease the slope, in contrast to the commonly assumed blue-shifting of the photon spectrum.
We make the case that a minimal degree of modelling of quark-gluon plasma expansion provides a reasonable mapping between  the slope of the photon energy spectrum and the maximum temperature of the plasma, as long as the slope is extracted at a reasonably high photon energy and thermal photons can be isolated from non-thermal sources.
Validating our approach using state-of-the-art numerical calculations of photon production, we highlight the challenges of contamination from non-thermal sources.
\end{abstract}

\maketitle

\section{Introduction}

While most photons produced in ultrarelativistic collisions of nuclei are by-products of hadronic decays, a small but measurable number has a more exotic origin: they are emitted directly by quark-gluon plasma, the short-lived plasma of deconfined nuclear matter produced in high-energy heavy-ion collisions~\cite{Shuryak:1978ij,Gale:2009gc,David:2019wpt,Monnai:2022hfs}.
The low-energy ($\sim 1$~GeV) spectrum of these non-decay photons has a near exponential dependence~\cite{PHENIX:2014nkk,ALICE:2015xmh,STAR:2016use,PHENIX:2022rsx,PHENIX:2022qfp},  $\exp(-E/\Teff)$. The inverse slope $\Teff$ is often interpreted as an ``effective temperature'' for the plasma. This interpretation of $\Teff$ originates from the approximate $\left[\exp(-E/T)\right]$ dependence of the photon emission rate, for photons of energy $E$ emitted by a plasma of temperature $T$.

A large body of work indicates that relativistic fluid dynamics can describe the spacetime evolution of quark-gluon plasma produced in nuclear collisions~\cite{Gale:2013da,Heinz:2013th,Shuryak:2014zxa,DerradideSouza:2015kpt}. While the plasma as a whole is globally out of equilibrium, the evidence suggests that one can treat it to a good approximation as being locally close to equilibrium.
This conclusion is largely based on the success of hydrodynamics-based models in describing a wide range of hadronic measurements from the Relativistic Heavy Ion Collider (RHIC) and the Large Hadron Collider (LHC)~\cite{Gale:2013da,Heinz:2013th,Busza:2018rrf,Heinz:2024jwu}.
In this picture, each space-time point in the plasma can be assigned a temperature; this temperature profile is inhomogeneous in space and rapidly varying in time.
Because of the complexity of this temperature profile,  it has long been understood that the exponential inverse slope $\Teff$ of the photon energy spectrum must represent, at best, a convoluted average over the plasma's temperature profile~\cite{Shuryak:1978ij,McLerran:1984ay}.
The interpretation of $\Teff$ is further complicated by the plasma's rapid expansion, which leads to a Doppler shift for the produced photons.
Finally, the exact rate of photon production by nuclear plasma is highly non-trivial, even assuming a thermalized system~\cite{Kapusta:1991qp, Aurenche:1998nw, Arnold:2001ms,Turbide:2003si,Liu:2006imd,Dusling:2009ej,Ghiglieri:2013gia,Gale:2014dfa,Lee:2014pwa,Holt:2015cda,Holt:2020mwf,Hidaka:2015ima,Ghiglieri:2016tvj,Ce:2022fot}:
while the $\left[\exp(-E/T)\right]$ suppression remains the dominant feature of the rate, there are well-known corrections~\cite{Kajantie:1981wg}.

All the above effects, and more\footnote{For example, modern numerical simulations include the effect of dissipation in the hydrodynamics description of the fluid, and at least partly account for the corresponding non-equilibrium effects in the photon emission rates~\cite{Shen:2013vja,Paquet:2015lta,Garcia-Montero:2019kjk}.}, can be included and studied in numerical simulations: hydrodynamics-based models provide detailed temperature and flow velocity profiles for the plasma, which can be combined with thermal photon production rates~\cite{Chaudhuri:2011up,Shen:2013vja,vanHees:2011vb,Chatterjee:2012dn,Paquet:2015lta,Kim:2016ylr,Dasgupta:2018pjm,Garcia-Montero:2019kjk,Monnai:2022hfs,Gale:2021emg}. 
The energy spectrum of photons found in these simulations is indeed near-exponential, and one can study numerically the relation between the slope of the photon spectrum and the simulation's actual temperature profile~\cite{vanHees:2011vb,Shen:2013vja,Garcia-Montero:2019kjk,Massen:2024pnj,Du:2025dot}. A similar exercise can be done with virtual photons (lepton pairs)~\cite{Rapp:2014hha,Churchill:2023zkk,Churchill:2023vpt,Du:2024pbd,Wu:2025iix}.

In this work, we approach the problem from a different angle.
We focus on head-on collisions of large nuclei (central heavy-ion collisions), which produce quark-gluon plasma with an approximate cylindrical symmetry in the plane transverse to the collision axis. 
These central collisions have the highest achievable nuclear matter density in collider experiments, increasing the plausibility of the local equilibrium assumption.
Moreover, the larger number of nucleons participating in the formation of quark-gluon plasma reduces event-by-event fluctuations in the temperature profile of the plasma. 
This allows for a more meaningful definition of an ``initial temperature profile'', whose maximum is approximately at the center of the nuclear overlap region.
We put forward a simpler core model to understand the photon energy spectrum and its inverse slope, and their relation to the plasma's maximum temperature.
We validate this model with numerical calculations.
We make the case that our approach provides a pragmatic compromise between accuracy and complexity to understand the photon spectrum.

As a result of our study, we provide new insights into the role of the plasma's transverse Doppler shift on the slope of the photon spectrum. We show in Section~\ref{sec:effect_of_transverse_flow} how the Doppler shift increases $\Teff$ at low photon energies and decreases it at higher photon energies, with a threshold around 2--3~GeV. The mapping between the inverse slope and the maximum temperature of the plasma is derived in Section~\ref{sec:photon_spectrum_no_doppler} with additional validation in Section~\ref{sec:comparison_with_calcs_and_data}. The role of non-thermal sources and their contribution to $\Teff$ is discussed in Section~\ref{sec:comparison_with_calcs_and_data}, where the approach is also compared with experimental data.

\section{Thermal photon energy spectrum at midrapidity}

\label{sec:thermal_photons_midrap}

Photon production in collisions of nuclei is described in terms of their momentum perpendicular and parallel to the collision axis. For a photon of momentum $K^\mu$, the rapidity $y_M=\ln[(K^t+K^z)/(K^t-K^z)]/2$ is used as hyperbolic angle to characterize the momentum along the beam axis, with $y_M=0$ perpendicular to this axis. Cylindrical coordinates are used in the direction orthogonal to the collision axis: $k_T$ and $\phi$. 

If we focus on photons produced by a locally thermalized plasma\footnote{If viscous hydrodynamics is used and the plasma deviates from local equilibrium, the rate in Eq.~\ref{eq:thermal_photons} depends on additional degrees of freedom, like the shear tensor and the bulk pressure, which must be mapped to viscous corrections of the photon emission rate. See e.g. Refs.~\cite{Baier:1997xc,Schenke:2006yp,Dusling:2009bc,Dion:2011pp,Shen:2014nfa,Hauksson:2017udm,Liu:2017fib,Kasmaei:2019ofu}.}, the $\phi$-averaged spectrum is given by~\cite{Kapusta:2006pm}
\begin{equation}
	 \frac{1}{2\pi k_T} \frac{d N}{dk_T dy_M} =  \int_0^{2\pi} \frac{d \phi}{2\pi} \int 
	 d\tau \tau d^2 x d\eta_s \left[ k \frac{d^3 \Gamma_{\gamma}(K\cdot u,T)}{d^3 k} \right]
	 \label{eq:thermal_photons}
\end{equation}
where $k d^3 \Gamma_{\gamma}(K\cdot u,T)/d^3 k$ is the rate of production of photons per spacetime volume~\cite{Kapusta:1991qp, Aurenche:1998nw, Arnold:2001ms,Turbide:2003si,Liu:2006imd,Dusling:2009ej,Ghiglieri:2013gia,Gale:2014dfa,Lee:2014pwa,Holt:2015cda,Holt:2020mwf,Kapusta:1991qp,Aurenche:1998nw,Arnold:2001ms,Ghiglieri:2013gia,Lee:2014pwa,Gale:2014dfa,Hidaka:2015ima,Ghiglieri:2016tvj,Ce:2022fot}. The temperature and flow velocity profile of the plasma, in space and time, are given respectively by $T(\tau,x,y,\eta_s)$ and  $u^\mu(\tau,x,y,\eta_s)$, expressed as a function of the hyperbolic spatial coordinates $\tau^2=t^2-z^2$ and $\tanh \eta_s =z/t$.

In symmetric collisions of large nuclei, the spacetime region around $\eta_s=0$ is approximately invariant under longitudinal boosts~\cite{BRAHMS:2004adc}.
In this Bjorken longitudinal boost-invariant limit~\cite{Bjorken:1982qr}, the temperature and transverse flow velocity are independent of $\eta_s$, and the $\eta_s$-component of the flow velocity can be neglected.
Photons produced at a given \emph{momentum} rapidity $y_M$ originate from a narrow window in \emph{spatial} rapidity $\eta_s$ around $\eta_s \sim y_M$~\cite{McLerran:1984ay}. 
For photons produced with negligible longitudinal momentum ($y_M=0$),
one can, to very good accuracy, approximate the plasma as boost-invariant.
Writing $K\cdot u=k_T \left[\cosh(\eta_s) \sqrt{1+u_\perp^2}-u_\perp \cos(\phi) \right]$ and using the fact that the dominant momentum dependence of the photon rate is $\exp(-K\cdot u/T)$ (see Appendix~\ref{sec:appendix_rate_exp}), integration over $\eta_s$ and $\phi$ yields~\cite{Kajantie:1986cu}:
\begin{multline}
	\frac{1}{2\pi k_T} \left.\frac{d N}{d k_T dy_M}\right|_{y_M=0} \!\!\!\!\!\!\!\!\! \approx \sqrt{2 \pi }  \int d\tau \tau d^2 x  e^{-\frac{k_T u_\perp}{T}} I_0\left( \frac{k_T u_\perp}{T} \right)  \\ \sqrt{\frac{T}{k_T \sqrt{1+u_\perp^2}}}  \left[ k \frac{d^3 \Gamma_{\gamma}(k_T (\sqrt{1+u_\perp^2}-u_\perp),T)}{d^3 k} \right]
\label{eq:spectra_with_uperp}
\end{multline}
where $u_\perp$ and $T$ depend on $\tau$, $x$ and $y$. 
The transverse proper velocity $u_\perp$ is related to the transverse three-velocity  $v_\perp$ by $u_\perp=v_\perp/\sqrt{1-v^2_\perp}$.
The function $I_0(a)$ is the modified Bessel function of the first kind; the product $e^{-a} I_0(a)$ is close to $1/(1+a)$ for $a\lesssim 5$ and $1/\sqrt{2\pi a}$ for $a \gg 1$. 

At $y_M= 0$, the transverse momentum $k_T$ of photons corresponds to their energy $E$, and there is no Jacobian:
\begin{equation*}
\frac{1}{2\pi E} \left.\frac{d N}{d E dy_M}\right|_{y_M=0} =\frac{1}{2\pi k_T} \left.\frac{d N}{d k_T dy_M}\right|_{y_M=0}
\end{equation*}
Hence, Eq.~\ref{eq:spectra_with_uperp} is truly the energy spectrum of photons. 
In this work, we will only focus on photons produced at midrapidity ($y_M= 0$), and hence will use $E$ and $k_T$ interchangeably, with a preference for $E$.

\section{Temperature, velocity and volume profile of the plasma}

\label{sec:effect_of_transverse_flow}

Given the symmetry of head-on heavy-ion collisions in the plane transverse to the collision axis, we approximate the initial temperature distribution of the plasma as radially symmetric. On an event-by-event basis, this symmetry is broken even in head-on collisions by fluctuations from the quantum nature of nuclear collisions~\cite{Dion:2011pp,Chatterjee:2012dn}.
The width of the initial temperature profile should be approximately the same as the radius of the colliding nuclei, $\sigma_0 \approx 5$--$10$~fm for heavy ions such as gold or lead. We use a Gaussian temperature distribution of width $\sigma_{0}$ and maximum temperature $\Tmax$:
\begin{align}
T(\tau_0,r)=\Tmax e^{-r^2/(2\sigma_{0}^2)}
\label{eq:T0_profile_gaussian}
\end{align}
with $r$ the radial distance in the plane transverse to the collision axis.

\begin{figure}
	\centering
	\includegraphics[width=0.99\linewidth]{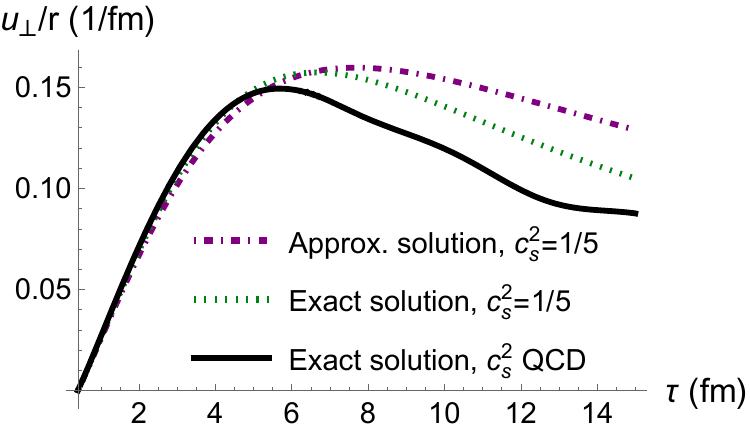}
	\caption{Ratio of the radial transverse flow velocity $u_\perp=u^r$ to the transverse radius $r$ in the $r\to 0$ limit for ideal transversely-cylindrically-symmetric boost-invariant hydrodynamics with a Gaussian radial initial temperature profile set at time $\tau_0=0.4$~fm with a maximum initial temperature of $\Tmax=0.55$~GeV and a width of $\sigma_0=5$~fm. The exact ideal hydrodynamics solution is shown for two equations of state: the solid line is with the QCD equation of state~\cite{Borsanyi:2013bia,Bernhard:2018hnz,eos_code}, while the dotted line is for an equation of state with a constant speed of sound squared $c_s^2=1/5$. The result is compared with the approximate Eq.~\ref{eq:uperp} (divided by $r$). The result is only shown for $r \to 0$ since the ratio $u_\perp/r$ is only weakly dependent on $r$.}
	\label{fig:flow_cs2_dep_and_approx}
\end{figure}

For $\tau > \tau_0$, we describe the evolution of the plasma's temperature and flow velocity profile with inviscid relativistic fluid dynamics. 
An approximate expression for the transverse proper flow velocity is\footnote{
Starting with the inviscid hydrodynamics equations for a Gaussian initial temperature profile, no initial transverse velocity and a constant speed of sound, a solution can be found for $\tau \ll \sigma_0$, yielding Eq.~\ref{eq:uperp} without the $1/(1 + \tau^2/(2 \sigma_0^2))$. The latter factor can be identified by comparisons with numerical solutions and dimensional analysis.}
\begin{equation}
u_\perp(\tau,r)\approx \frac{r}{\sigma_0} \frac{ \left(\tau - \left(\frac{\tau}{\tau_0}\right)^{c_s^2} \tau_0 \right)}{(1 - c_s^2) \sigma_0 (1 + \tau^2/(2 \sigma_0^2))} 
\label{eq:uperp}
\end{equation}
where $c_s^2$ is the speed of sound squared of the plasma (in units of the speed of light), and we assumed that the \emph{transverse} flow velocity is negligible at time $\tau_0$. %
We show a comparison of Eq.~\ref{eq:uperp} with exact numerical solutions in Figure~\ref{fig:flow_cs2_dep_and_approx}; we see that Eq.~\ref{eq:uperp}  does indeed capture well the general dependence of the transverse flow velocity on the time $\tau$.

The effect of the transverse flow velocity on the photon spectrum (Eqs.~\ref{eq:thermal_photons} and ~\ref{eq:spectra_with_uperp}) depends on three factors: (i) the magnitude of the transverse flow, (ii) the effect of transverse flow on photons of different energies, and (iii) the origin of the photons that dominates the spectrum at different energies.

\subsection{Magnitude of transverse flow}

\label{sec:transverse_flow_vs_r_and_tau}

\begin{figure}
	\centering
	\includegraphics[width=0.8\linewidth]{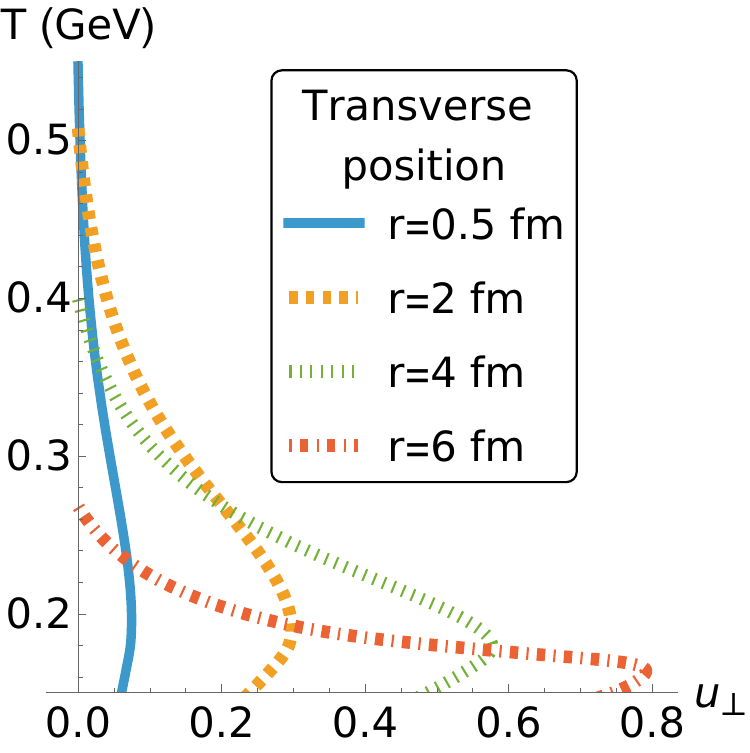}
	\caption{
		For different fixed values of the radius $r$, each line shows the trajectory in temperature $T(\tau,r)$ and transverse flow velocity $u_\perp(\tau,r)$ as $\tau$ is varied. 
		The highest temperature regions of the plasma (top part of figure) tend to have small transverse flow $u_\perp$, while large values of $u_\perp$ (right part of figure) are mainly achieved at low temperatures.
		 Obtained from a numerical solution of  ideal transversely-cylindrically-symmetric boost-invariant hydrodynamics with a Gaussian radial initial temperature profile set at time $\tau_0=0.4$~fm with a maximum initial temperature of $\Tmax=0.55$~GeV and a width of $\sigma_0=5$~fm and the QCD equation of state~\cite{Borsanyi:2013bia,Bernhard:2018hnz,eos_code}. 
	}
	\label{fig:T_uperp_trajectories}
\end{figure}

Equation~\ref{eq:uperp} provides an answer for the first factor: transverse flow is small when $r, (\tau-\tau_0) \ll \sigma_0$, which are both regions where the plasma's temperature is large. 
This is well known from numerical hydrodynamic simulations.
For small $\tau$, the plasma is hot because it has not cooled down significantly yet. For small $r$, the plasma is hot because a larger density of nuclear material, integrated along the collision axis, is deposited on impact.
This is not to say that the correlation between temperature and the transverse flow is simple.
 However, given the overall setup of the problem (central collisions), high-temperature regions of the plasma tend to have relatively small values of $u_\perp$. This is shown in Figure~\ref{fig:T_uperp_trajectories} using the results of numerical simulations.

\subsection{Local effect of Doppler shift}

\label{sec:local_doppler}

If one assumes that the thermal emission rate is approximately exponential --- $k d^3 \Gamma_\gamma/d^3 k \propto \exp(-E/T)$ (see Appendix~\ref{sec:appendix_rate_exp}) --- and uses $u_\perp=v_\perp/\sqrt{1-v^2_\perp}$, then
\begin{multline}
	k \frac{d^3 \Gamma_{\gamma}(E (\sqrt{1+u_\perp^2}-u_\perp),T)}{d^3 k} \\
	\propto \exp\left(-\frac{E (\sqrt{1+u_\perp^2}-u_\perp)}{T}\right) \\
	= \exp\left(-\frac{E}{T \sqrt{\frac{1+v_\perp}{1-v_\perp}}}\right)
\end{multline}
with the temperature multiplied by the oft-quoted factor of
\begin{align*}
	\sqrt{\frac{1+v_\perp}{1-v_\perp}},
\end{align*}
which \emph{partially} captures the effect of the Doppler shift on a \emph{single localized volume of the plasma}. Indeed,
even with the approximation above, the integrand of Eq.~\ref{eq:spectra_with_uperp} is proportional to
\begin{align*}
	\exp\left(-\frac{E}{T(X) \sqrt{\frac{1+v_\perp(X)}{1-v_\perp(X)}}}\right) \frac{\sqrt[4]{1-v_\perp^2(X)}}{
		\left(1+\frac{E}{T(X)}\frac{ v_\perp(X)}{\sqrt{1-v_\perp(X)^2}}\right)}
\end{align*}
up to additional constants and factors of temperature $T(X)$ and photon energy $E$. The dependence on spacetime ($X$) of the transverse flow and the temperature was restored to emphasize that this quantity is defined locally, and is not necessarily a meaningful quantity after spacetime integration. 

The local Doppler shift of photons can be estimated from the integrand of Eq.~\ref{eq:spectra_with_uperp}.
Assuming again an exponential photon emission rate, we can expand the integrand at $u_\perp=0$ and factor out the $u_\perp$-dependent part, yielding
\begin{equation}
\left[1+\frac{u_\perp^2}{4} \left[ \frac{E}{T} \left( \frac{E}{T} -2 \right) -1 \right]\right] \times (\textrm{integrand at $u_\perp=0$}) .
\label{eq:uperp_effect}
\end{equation}
Equation~\ref{eq:uperp_effect} summarizes that the effect of transverse flow can only be large for $E \gg T$. This is a well-known effect which has been discussed and quantified numerically in multiple earlier publications~\cite{vanHees:2011vb,Shen:2013vja,Paquet:2016ime}. 
This, however, does not imply that the effect of the Doppler shift is necessarily large in the photon spectrum for $E\gg T$, since this is purely a local effect, and the total spectrum is a convolution of such complicated local effects with variable temperatures and transverse flow velocities, as discussed in the next section.

\begin{figure}
	\centering
	\includegraphics[width=0.99\linewidth]{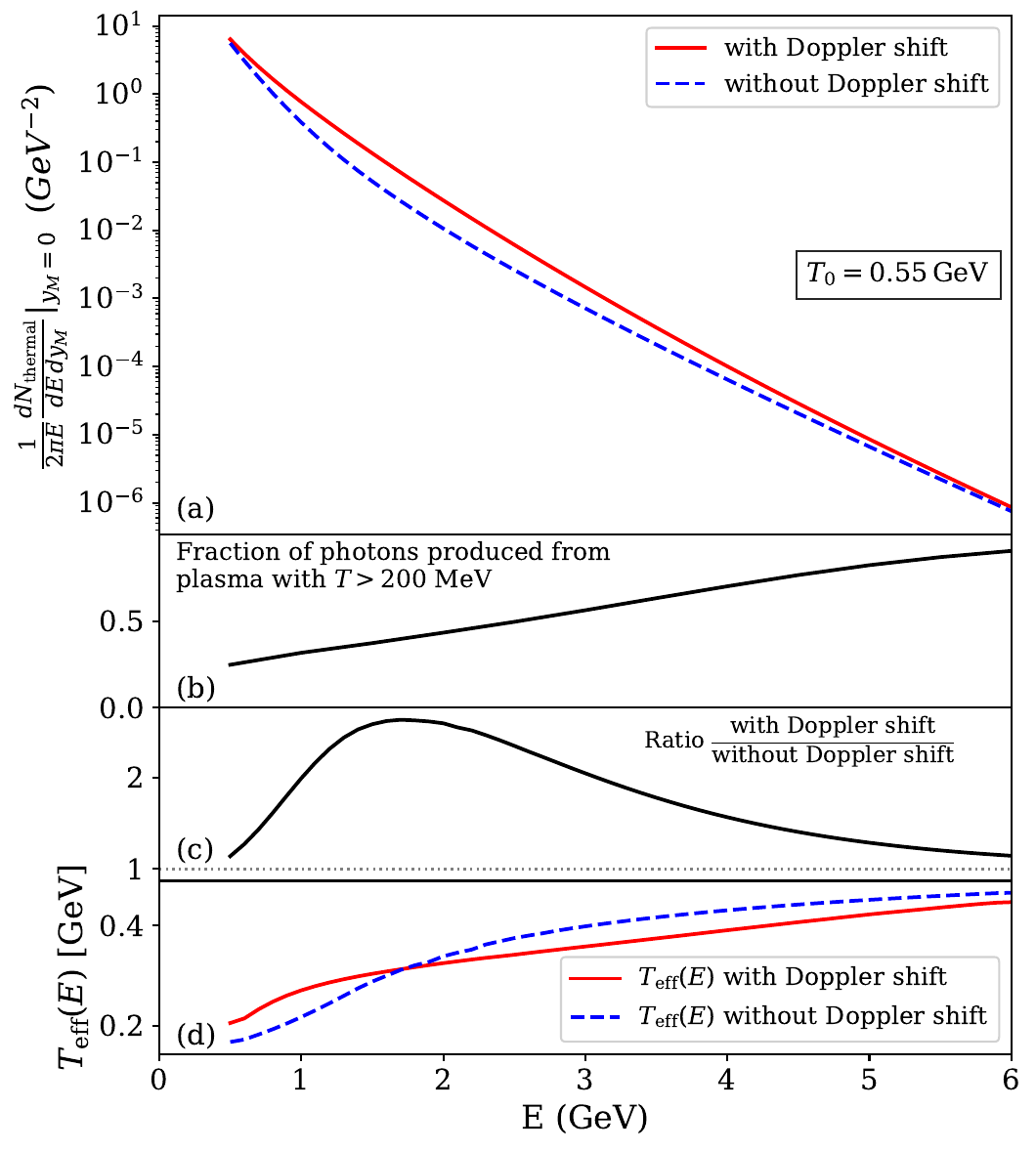}
	\caption{(a) Photon spectrum with (solid red line) and without (dashed blue line) the Doppler effect due to the transverse expansion $u_\perp$ in  Eq.~\ref{eq:spectra_with_uperp}. The calculation is for a temperature and flow velocity profile produced with  ideal transversely-cylindrically-symmetric boost-invariant hydrodynamics with a Gaussian radial initial temperature profile set at time $\tau_0=0.4$~fm with a maximum initial temperature of $\Tmax=0.55$~GeV and a width of $\sigma_0=5$~fm, and with a realistic QCD equation of state (same as Figure~\ref{fig:T_uperp_trajectories}). Photons are produced for temperatures above $T=0.12$~GeV. The photon emission rate from Ref.~\cite{Arnold:2001ms} is used. (b) Fraction of photons produced from regions of the plasma where the temperature is above $0.2$~GeV. 
	(c) Ratio of the two calculations from panel (a).
	(d) Inverse slope $\Teff$ from $\frac{1}{2\pi E} \left.\frac{d N_{\rm thermal}}{d E dy_M}\right|_{y_M=0} \propto \exp(-E/\Teff)$ as a function of $E$ for the two calculations from panel (a).
	}
	\label{fig:spectrum_with_without_flow}
\end{figure}

\begin{table*}[tb]
	\centering
	\begin{tabular}{|l|c|c|c||c|c|c||c|c|c|}
		\hline
		& \multicolumn{3}{|c||}{\textrm{$\Tmax$=0.35 GeV}} & \multicolumn{3}{|c||}{\textrm{$\Tmax$=0.55 GeV}} & \multicolumn{3}{|c|}{\textrm{$\Tmax$=0.75 GeV}} \\
		\hline
		& \multicolumn{2}{|c|}{\textrm{$\Teff$}} & & \multicolumn{2}{|c|}{\textrm{$\Teff$}} & & \multicolumn{2}{|c|}{\textrm{$\Teff$}} & \\
		\hline
		\textrm{$E$ range} & \textrm{w/ Doppler} & \textrm{w/o Doppler} & \textrm{rel. diff.} & \textrm{w/ Doppler} & \textrm{w/o Doppler} & \textrm{rel. diff.} & \textrm{w/ Doppler} & \textrm{w/o Doppler} & \textrm{rel. diff.} \\
		\textrm{[GeV]} & \textrm{[GeV]}  & \textrm{[GeV]}  &  & \textrm{[GeV]}  & \textrm{[GeV]}  &  & \textrm{[GeV]}  & \textrm{[GeV]}  &  \\
		\hline
		\textrm{full [0.5, 6.0]} & $0.257\pm0.002$ & $0.264\pm0.003$ & $-3\%$ & $0.357\pm0.004$ & $0.373\pm0.006$ & $-4\%$ & $0.446\pm0.005$ & $0.464\pm0.009$ & $-4\%$ \\
		\hline
		\textrm{low [0.5, 2.0]} & $0.210\pm0.002$ & $0.192\pm0.003$ & $+10\%$ & $0.281\pm0.004$ & $0.244\pm0.007$ & $+15\%$ & $0.338\pm0.007$ & $0.284\pm0.010$ & $+19\%$ \\
		\hline
		\textrm{mid [2.0, 4.0]} & $0.256\pm0.002$ & $0.274\pm0.002$ & $-6\%$ & $0.357\pm0.002$ & $0.394\pm0.003$ & $-9\%$ & $0.450\pm0.002$ & $0.495\pm0.004$ & $-9\%$ \\
		\hline
		\textrm{high [4.0, 6.0]} & $0.297\pm0.001$ & $0.304\pm0.001$ & $-2\%$ & $0.420\pm0.002$ & $0.450\pm0.001$ & $-7\%$ & $0.522\pm0.002$ & $0.578\pm0.002$ & $-10\%$ \\
		\hline
	\end{tabular}
	\caption{Extracted inverse slope $\Teff$ from $\frac{1}{2\pi E} \left.\frac{d N_{\rm thermal}}{d E dy_M}\right|_{y_M=0} \propto \exp(-E/\Teff)$ over different ranges of $E$, comparing Doppler-shifted vs non-shifted spectra, and relative difference between these $\Teff$. The photon spectra include only thermal photons produced from inviscid relativistic hydrodynamics with Gaussian initial conditions, as described in the caption of Figure~\ref{fig:spectrum_with_without_flow} and in the text. 
		The uncertainty on the inverse slope is the standard error from the linear regression; it is \emph{not} a theoretical uncertainty. 
		}
	\label{tab:Teff_ranges}
\end{table*}

\subsection{Global effect of the Doppler shift on the photon spectrum}
It is known from numerical simulations~\cite{Shen:2013vja,vanHees:2011vb,Paquet:2016ulk, Paquet:2016ime,Monnai:2022hfs,Gale:2021emg} that photons produced at high temperatures dominate the photon energy spectrum at high energy ($E \gtrsim 2$--$3$~GeV), and photons produced at low temperatures dominate at low energy ($E \lesssim 2$--$3$~GeV).
To illustrate this point, we compute thermal photons using cylindrically-symmetric boost-invariant inviscid hydrodynamics profile evaluated numerically with a realistic QCD equation of state~\cite{Borsanyi:2013bia,Bernhard:2018hnz,eos_code} and a Gaussian initial condition in the transverse direction (Eq.~\ref{eq:T0_profile_gaussian}). The photon spectrum is calculated with and without the effect of the Doppler shift from the transverse flow velocity (that is, $u_\perp$ is set to zero in Eqs.~\ref{eq:thermal_photons} and ~\ref{eq:spectra_with_uperp}); we  evaluate it numerically using the photon emission rate from Ref.~\cite{Arnold:2001ms} with three quark flavors and the strong coupling set to $g_s=2$ ($\alpha_s=0.32$).
For the hydrodynamics, we set the initial time $\tau_0=0.4$~fm with a maximum initial temperature of $\Tmax=0.55$~GeV and a width of $\sigma_0=5$~fm.\footnote{In this section and the next, we show results for a single value of $\sigma_0$ and $\tau_0$.	We verified that the paper's conclusions hold for different values of these parameters. Similarly, we verified that our conclusions hold if the photon emission rate from Ref.~\cite{Arnold:2001ms} is replaced at lower temperature ($T<0.18$~GeV) by a more realistic rate evaluated with hadronic degrees of freedom~\cite{Turbide:2003si,Liu:2006imd,Heffernan:2014mla,Holt:2015cda}.}
The energy spectrum is shown in Figure~\ref{fig:spectrum_with_without_flow}(a).
In panel (b), we show the fraction of photons produced at high temperature, defined here to be $T>200$~MeV. 
We see that most high-energy photons are indeed produced at high temperature.
As discussed above, these high-temperature regions generally have small transverse flow velocities, and consequently, are not expected to have a large effect of the Doppler shift.
This is clearly visible in panel (a), as well as in panel (c), which shows the ratio of the photon spectrum calculated with the Doppler shift to the calculation without the Doppler shift.
Photons produced in the lower temperature regions of the plasma (large $r$ or $\tau$), where the transverse flow is large, dominate the lower energy part of the spectrum; however, as explained previously (Eq.~\ref{eq:uperp_effect}), the effect of transverse flow is suppressed for small $E/T$.
In consequence, the effect of the transverse Doppler shift on the photon spectrum is highest at \emph{intermediate} thermal photon energies. This intermediate energy range is $E \approx 1.5$--$3$~GeV for the simulations used in Figure~\ref{fig:spectrum_with_without_flow}, but could vary based on the details of the simulation (for example, center-of-mass energy of the collision, effect of viscosity, type of initial conditions).

\subsection{Global effect of the Doppler shift on the inverse slope $\Teff$}

For the independent reasons discussed above, the effect on photons from the plasma's transverse expansion is suppressed for both the low and high energy range of the photon spectrum.
To understand the consequence on the inverse slope, we plot it as a function of the photon energy $E$ in panel (d) of 
Figure~\ref{fig:spectrum_with_without_flow}.

We see that the Doppler shift does not necessarily increase the inverse slope $\Teff$, as is typically assumed based on the Doppler shift's local effect (Section~\ref{sec:local_doppler}). In fact, there is a significant photon energy range where the Doppler shift decreases the inverse slope.
Indeed, if the effect of the Doppler shift is small at both low and high $E$, the inverse slopes must converge at these points. This means that if there is any range in energy $E$ where the Doppler shift increases the inverse slope, it has to be compensated in another range in $E$ by the Doppler shift decreasing the inverse slope. The energy at which the effect of the Doppler shift is maximal \emph{on the spectrum} is also the energy where it has \emph{zero} effect on the inverse slope, meaning that at this point, the inverse slopes with and without Doppler shift are strictly equal:
\begin{align}
\frac{d \ln \left( \frac{d N_{w}(E)}{d N_{w/o}(E)} \right)}{d E}&=\frac{d \ln d N_{w} }{d E} - \frac{d \ln d N_{w/o} }{d E} \nonumber \\
& =-\frac{1}{\Teff^{(w)}}+\frac{1}{\Teff^{(w/o)}}
\label{eq:dlog_ratio}
\end{align}
where we used $d N_{w}(E)$ and $d N_{w/o}(E)$ as shorthand notation for the thermal photon spectrum with and without transverse Doppler shift, and where we used $\Teff^{(w)}$ for the inverse slope with Doppler shift included and $\Teff^{(w/o)}$ for the inverse slope without. In summary, if one computes the ratio of the spectrum with and without Doppler shift, the sign of the derivative of this ratio indicates whether the Doppler shift increases the inverse slope or decreases it.

Experiments extract the inverse slope in relatively wide energy  $E$ bins (or transverse momentum $k_T$ bins). To match more accurately this procedure, we show in Table~\ref{tab:Teff_ranges} the inverse slope $\Teff$ for four different bins of energy, for the same hydrodynamic calculation as used in Figure~\ref{fig:spectrum_with_without_flow}. To explore the dependence of the result on the initial temperature of the plasma, we perform this procedure for three values of $\Tmax$: $0.35$~GeV, $0.55$~GeV (used in Figure~\ref{fig:spectrum_with_without_flow}) and $0.75$~GeV. Changing the initial temperature serves as a proxy for changing the center-of-mass energy of the nuclear collision. We see that the Doppler shift indeed \emph{reduces} the inverse slope extracted with $E \in [2,4]$ and $E \in [4,6]$. The effect of the Doppler shift on the inverse slope is less than $10$\%. On the other hand, if the inverse slope is extracted from the lower end of the photon spectrum ($E \in [0.5,2]$), the Doppler shift does increase the inverse slope, by $10$\% to $20$\%.

We verified that similar conclusions can be obtained from more complete calculations of the photon spectrum from an earlier publication~\cite{Paquet:2015lta}. These results are presented in Appendix~\ref{sec:Teff_prc}.

\subsection{Assessment of the effect of the transverse Doppler shift}

In absolute magnitude, the transverse Doppler shift is a significant effect on the photon energy spectrum at intermediate energies. On the other hand, as discussed above, the effect on the inverse slope $\Teff$ is much more subtle. Depending on the energy range in which the inverse slope is extracted, we estimated the effect of the Doppler shift to be $-10$\% to $+20$\%. While not negligible, this suggests that one can still obtain a qualitative and even quantitative understanding of the inverse slope $\Teff$ while neglecting the effect of the transverse Doppler shift, which allows for very significant simplifications. This is the approach we use in the rest of this work.

\section{Photon spectrum without transverse Doppler shift, and the interpretation of the inverse slope}

\label{sec:photon_spectrum_no_doppler}

Setting $u_\perp$ to zero in Eq.~\ref{eq:spectra_with_uperp}, the photon spectrum can be written as
\begin{multline}
	\frac{1}{2\pi E} \left.\frac{d N}{d E dy_M}\right|_{y_M=0} \!\!\!\!
	\approx \int_{T_f}^{\Tmax} \!\!  d T \frac{d V_\perp}{d T} \sqrt{\frac{2\pi T}{E}} \\ \times \left[ k \frac{d^3 \Gamma_{\gamma}(E,T)}{d^3 k} \right] \\
	\label{eq:spectra_with_dV_dT}
\end{multline}
where we used the transverse spacetime volume of plasma per unit spatial rapidity at temperature $T$ given by~\cite{Shuryak:1978ij} 
\begin{align}
\frac{d V_\perp}{d T} \!=\!\int \!\! d\tau \tau d^2 r \delta(T - T(\tau,r) ) .
\label{eq:transverse_volume_def}
\end{align}
The maximum of the integration range is $\Tmax$, the maximum temperature of the plasma, found at time $\tau_0$. The lower bound is the cutoff $T_f$. Guidance for the choice of $T_f$ can be found in studies of the later stage of nuclear collisions, where a transition from fluid dynamics to hadronic transport is necessary given the lower density of nuclear matter. Based on Ref.~\cite{Gotz:2021dco}, we can estimate $T_f \approx 120$--$140$~MeV, although the spectrum of higher-energy photons does not depend significantly on this value.

The transverse volume $d V_\perp/d T$ is only known analytically for simple inviscid hydrodynamic solutions, like for the Gubser solution~\cite{Paquet:2023bdx}.
It can be found for the Bjorken solution to hydrodynamics as well, which we compute below.
Assuming a Gaussian temperature profile (Eq.~\ref{eq:T0_profile_gaussian}), and assuming a Bjorken expansion with a constant speed of sound $c_s^2$ --- $T(\tau,r)=T(\tau_0,r) (\tau_0/\tau)^{c_s^2}$ --- we find
\begin{align}
	\frac{d V_\perp}{ d T} \approx \frac{\pi  \sigma_0^2 \tau_0^2 \left( \left(\frac{\Tmax}{T}\right)^{2 c_s^{-2}}-1\right)}{T}
	\label{eq:dVdT_gaussian_cs2}
\end{align}

A more general formula (Eq.~\ref{eq:dVdT_var_cs2}) can be derived in the case where the speed of sound is temperature dependent, as is that of QCD plasmas. However, as discussed in Appendix~\ref{sec:appendix_dVdT}, this seeming improvement brings complexity in return for limited benefits, except for collisions of nuclei at lower energies than the ones considered in this manuscript. This is due to accidental cancellations of errors between the equation of state and the transverse expansion, which often makes $c_s^2=1/3$ a good approximation, even when describing the QCD plasma formed in ultrarelativistic nuclear collisions. 
We thus assume a constant $c_s^2$ for the rest of this work. 

The accidental cancellation of errors discussed in Appendix~\ref{sec:appendix_dVdT} is key to the surprising accuracy of Bjorken hydrodynamics to estimate the transverse volume of the plasma.
The Gubser solution does not have this accidental cancellation --- it has overly rapid expansion compared to a QCD plasma from assuming $c_s^2=1/3$, while also having transverse flow velocity. This leads to a cooling that is significantly faster than in heavy-ion collisions.

\begin{figure}[t]
	\centering
	\includegraphics[width=0.99\linewidth]{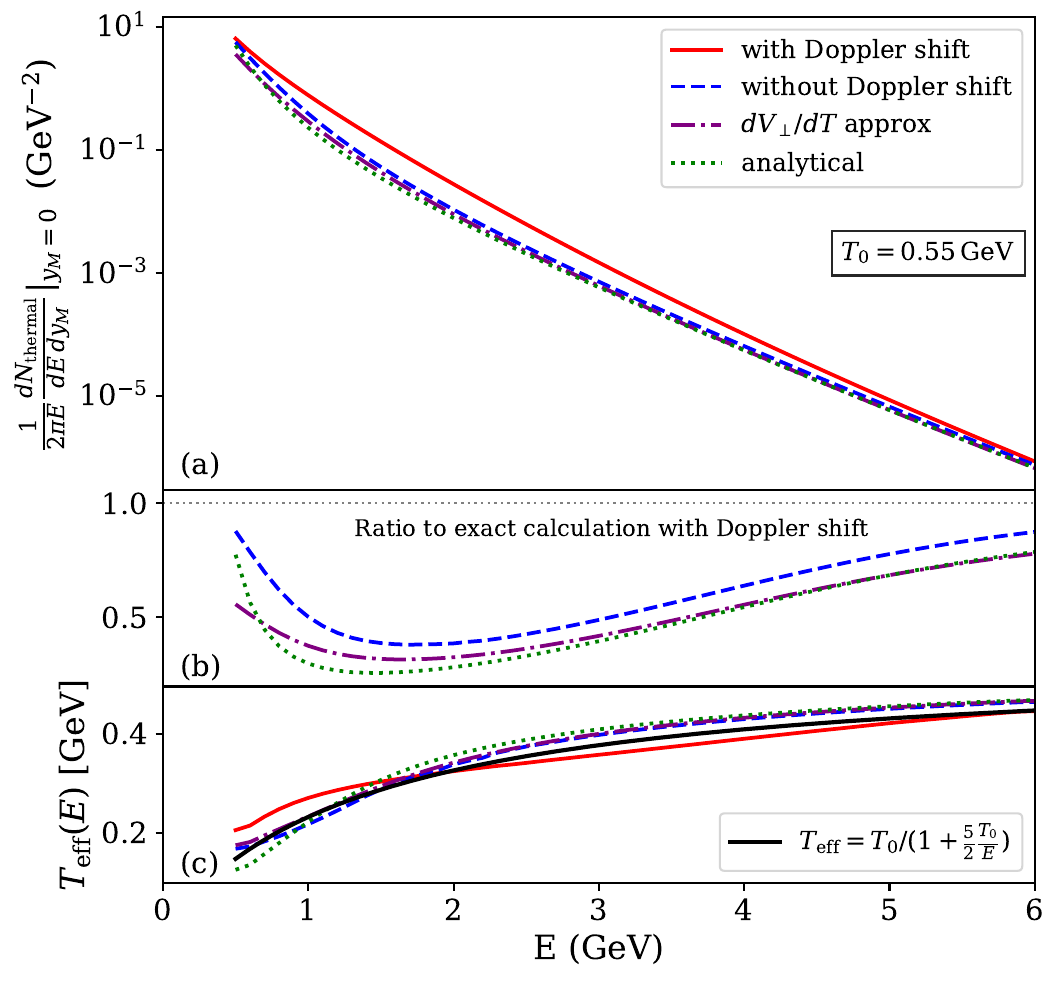}
	\caption{(a) Exact thermal photon spectrum as a function of the photon energy $E$ obtained for the same ideal hydrodynamics evolution as used in Figure~\ref{fig:spectrum_with_without_flow} (solid red line), compared to the result without transverse Doppler shift (blue dashed line), and compared to two additional approximations: Eq.~\ref{eq:spectra_with_dV_dT} with $d V_\perp/d T$ from Eq.~\ref{eq:dVdT_gaussian_cs2} (purple dash-dotted line), and the analytical approximation from Eq.~\ref{eq:spectra_with_dV_dT_conformal}.
		(b) Ratio of the calculations from panel (a) to the exact calculation.
		(c) Inverse slope $\Teff$ from $\frac{1}{2\pi E} \left.\frac{d N_{\rm thermal}}{d E dy_M}\right|_{y_M=0} \propto \exp(-E/\Teff)$ as a function of $E$ for the four calculations from panel (a). For all panels, the red line and the dashed blue line are the same as in Figure~\ref{fig:spectrum_with_without_flow}.
	}
	\label{fig:spectrum_approxs}
\end{figure}

\begin{table*}[!tb]
	\centering
	\begin{tabular}{|l|c|c|c||c|c|c||c|c|c|}
		\hline
		& \multicolumn{3}{|c||}{\textrm{$\Tmax$=0.35 GeV}} & \multicolumn{3}{|c||}{\textrm{$\Tmax$=0.55 GeV}} & \multicolumn{3}{|c|}{\textrm{$\Tmax$=0.75 GeV}} \\
		\hline
		\textrm{$E$ range} & \textrm{$\Teff[E_1,E_2]$} & \textrm{$\Tmax$ estimate} & \textrm{error} & \textrm{$\Teff[E_1,E_2]$} & \textrm{$\Tmax$ estimate} & \textrm{error} & \textrm{$\Teff[E_1,E_2]$} & \textrm{$\Tmax$ estimate} & \textrm{error} \\
		\textrm{[GeV]} & \textrm{[GeV]}  & \textrm{[GeV]}  &  & \textrm{[GeV]}  & \textrm{[GeV]}  &  & \textrm{[GeV]}  & \textrm{[GeV]}  &  \\
		\hline
		\textrm{full [0.5, 6.0]} & $0.257\pm0.002$ & $0.362\pm0.004$ & $+3\%$ & $0.357\pm0.004$ & $0.598\pm0.011$ & $+9\%$ & $0.446\pm0.005$ & $0.899\pm0.020$ & $+20\%$ \\
		\hline
		\textrm{low [0.5, 2.0]} & $0.210\pm0.002$ & $0.408\pm0.008$ & $+17\%$ & $0.281\pm0.004$ & $0.801\pm0.033$ & $+46\%$ & $0.338\pm0.007$ & $1.543\pm0.146$ & $+106\%$ \\
		\hline
		\textrm{mid [2.0, 4.0]} & $0.256\pm0.002$ & $0.329\pm0.003$ & $-6\%$ & $0.357\pm0.002$ & $0.517\pm0.004$ & $-6\%$ & $0.450\pm0.002$ & $0.738\pm0.005$ & $-2\%$ \\
		\hline
		\textrm{high [4.0, 6.0]} & $0.297\pm0.001$ & $0.350\pm0.001$ & $+0\%$ & $0.420\pm0.002$ & $0.534\pm0.003$ & $-3\%$ & $0.522\pm0.002$ & $0.710\pm0.004$ & $-5\%$ \\
		\hline
	\end{tabular}
	\caption{Maximum temperature of the plasma $\Tmax$ compared with estimates obtained from Eq.~\ref{eq:T0_estimate}. The inverse slopes extracted in finite energy ranges are the exact calculations from Table~\ref{tab:Teff_ranges} (with Doppler shift), obtained from ideal hydrodynamic simulations as discussed in Table~\ref{tab:Teff_ranges}.}
	\label{tab:T0_estimates}
\end{table*}

Evaluating Eq.~\ref{eq:spectra_with_dV_dT} with a simplified expression for the transverse volume $d V_\perp/d T$ is a further approximation. 
We show in Appendix~\ref{sec:appendix_dVdT} that the Bjorken approximation captures well the temperature dependence of transverse volume compared to numerical calculations of $d V_\perp/d T$. In what follows, we use Eq.~\ref{eq:dVdT_gaussian_cs2} to estimate $d V_\perp/d T$.

We make use of the approximate $T^2 \exp(-E/T)$ temperature dependence of the thermal emission rate $k d^3 \Gamma_{\gamma}(K\cdot u,T)/d^3 k$~\cite{Arnold:2001ms} (see Appendix~\ref{sec:appendix_rate_exp}). %
Keeping only the dominant term in $(T_f/E)$, $(T_f/\Tmax)$ and $(\Tmax/E)$ after performing the integral in Eq.~\ref{eq:spectra_with_dV_dT} with Eq.~\ref{eq:dVdT_gaussian_cs2}, we find
\begin{multline}
\frac{1}{2\pi E} \left.\frac{d N}{d E dy_M}\right|_{y_M=0} \!\!\!\!\!
\approx \left[ \frac{\exp(E/\Tmax)}{\Tmax^2} k \frac{d^3 \Gamma_{\gamma}(E,\Tmax)}{d^3 k} \right]  \\  (2\pi)^{3/2} c_s^{-2} \sigma_0^2 \tau_0^2 \Tmax^2 e^{-E/\Tmax} \left(\frac{\Tmax}{E}\right)^{5/2}  \\
\label{eq:spectra_with_dV_dT_conformal}
\end{multline}
We compare the four different results in Figure~\ref{fig:spectrum_approxs}: the exact numerical result with Doppler shift, the numerical result without Doppler shift, the result of 
Eq.~\ref{eq:spectra_with_dV_dT} with Eq.~\ref{eq:dVdT_gaussian_cs2} for $d V_\perp/d T$, and finally Eq.~\ref{eq:spectra_with_dV_dT_conformal}. The approximations differ by first dropping the Doppler shift, then approximating $d V_\perp/d T$ with Eq.~\ref{eq:dVdT_gaussian_cs2}, then simplifying this last result into an analytical expression. From Figure~\ref{fig:spectrum_approxs}, we see that the most significant approximation is neglecting the Doppler shift.

The inverse slope is related to the derivative of the logarithm of Eq.~\ref{eq:spectra_with_dV_dT_conformal}, yielding
\begin{align}
-\frac{1}{\Teff}=&-\frac{1}{\Tmax}-\frac{5}{2} \frac{1}{E} \nonumber \\
&+\frac{d}{d E} \ln \left[ \frac{\exp(E/\Tmax)}{\Tmax^2} k \frac{d^3 \Gamma_{\gamma}(E,\Tmax)}{d^3 k} \right]
\label{eq:slope}
\end{align}

Neglecting the non-exponential contributions to the rate, the result is
\begin{align}
\Teff=\frac{\Tmax}{1+\frac{5}{2} \frac{\Tmax}{E}}
\label{eq:simple_slope}
\end{align}
The same result was obtained in Ref.~\cite{Paquet:2023bdx} using a Gubser temperature profile, suggesting a degree of independence of this result from the exact form of the temperature profile at early times.

The result from Eq.~\ref{eq:simple_slope} is plotted in Figure~\ref{fig:spectrum_approxs}(c) as the solid black line. It tracks generally well the exact inverse slope (solid red line). The inverse slopes calculated from the various approximations discussed above are also shown. All approximations are similar, and all clearly indicate a significant difference between $\Teff$ and $\Tmax$ that is well approximated by Eq.~\ref{eq:simple_slope}.

To validate the practical usefulness of Eq.~\ref{eq:simple_slope} in estimating the maximum temperature of the plasma, we compare it with the inverse slopes computed in finite energy ranges, as done in Table~\ref{tab:Teff_ranges}. In this case, Eq.~\ref{eq:simple_slope} can be evaluated at the logarithmic mean of the bin:
\begin{align}
E_{\rm eff} = \frac{E_2 - E_1}{\ln(E_2/E_1)}
\label{eq:Eeff}
\end{align}
for a bin extending from $E_1$ to $E_2$. 

For an inverse slope $\Teff[E_1,E_2]$ extracted in an energy range $E_1$ to $E_2$, our estimate for the maximum temperature is given by
\begin{align}
\boxed{\Tmax=\frac{\Teff[E_1,E_2]}{1-\frac{5}{2} \frac{\Teff[E_1,E_2]}{E_{\rm eff} }}=\frac{\Teff[E_1,E_2]}{1-\frac{5}{2} \frac{\Teff[E_1,E_2] \ln(E_2/E_1)}{E_2 - E_1}}}
\label{eq:T0_estimate}
\end{align}
We highlight two distinct features of Eq.~\ref{eq:T0_estimate}. The inverse slope $\Teff[E_1,E_2]$ only converges to the maximum temperature $\Tmax$ if the slope is extracted for very high photon energy.
Second, we see that Eq.~\ref{eq:T0_estimate} breaks down if the photon's effective energy is equal or lower than $(5/2) \Teff[E_1,E_2]$.

We validate Eq.~\ref{eq:T0_estimate} by comparing it with the inverse slopes extracted in the previous section (Table~\ref{tab:Teff_ranges}). The results are shown in Table~\ref{tab:T0_estimates}: using $\Teff$ extracted in four different photon energy ranges, we attempt to extract three different $\Tmax$. Using Eq.~\ref{eq:T0_estimate} with a slope extracted in the high $E$ range ($E \in [4,6]$~GeV) yields an excellent estimate of the maximum temperature of the plasma, within $5\%$ of the correct value for all three $\Tmax$. A similar result is found even if the inverse slope is extracted in an $E \in [2,4]$~GeV energy range. However, Eq.~\ref{eq:T0_estimate} fails dramatically when the inverse slope is extracted in the lower end of the photon energy spectrum, in part due to large $\Tmax/E$ corrections in that limit. From these results, we see that a safer criterion of the $[E_1,E_2]$ bin where Eq.~\ref{eq:T0_estimate} can be used is approximately
\begin{align}
\boxed{E_{\rm eff}\equiv\frac{E_2 - E_1}{\ln(E_2/E_1)} \gtrsim  5 \Teff[E_1,E_2]}
\label{eq:Eeff_estimate}
\end{align}

Recall that the low-energy spectrum is dominated by photons produced at low temperatures, and hence late times. Estimating analytically the transverse volume $d V_\perp/d T$ (Eq.~\ref{eq:transverse_volume_def}) at late times is very challenging, and the cancellation of errors between the speed of sound and the transverse expansion discussed above (and in Appendix~\ref{sec:appendix_dVdT}) only goes so far. 
Moreover, if $E \sim \Tmax$, it is not accurate anymore to assume that $E$ is much larger than all other energy scales in the problem. Hence, for low photon energy, many of the approximations that lead to Eqs.~\ref{eq:spectra_with_dV_dT_conformal}-\ref{eq:simple_slope} become less accurate, or simply break down. This is not surprising: it should be simpler to constrain the plasma's maximum temperature from photons emitted at earlier times (higher energy photons). Hence, our results do not provide an interpretation of the inverse slope in all photon energy ranges, but Table~\ref{tab:T0_estimates} does confirm that it can provide a solid estimate of the plasma's maximum temperature if the inverse slope is extracted from the photon spectrum at reasonably large energies (Eq.~\ref{eq:Eeff_estimate}).

Finally, we note that Eq.~\ref{eq:T0_estimate} provides an estimate for the propagation of the uncertainty on $\Teff$ to $\Tmax$:
\begin{align}
\frac{\Delta \Tmax}{\Tmax} &\approx \frac{1}{1- \frac{5}{2} \frac{\Teff}{(E_2-E_1)} \ln[\frac{E_2}{E_1}]} \frac{\Delta \Teff}{\Teff} \nonumber \\
&=  \frac{1}{1- \frac{5}{2} \frac{\Teff}{E_{\rm eff}}} \frac{\Delta \Teff}{\Teff}
\label{eq:T0_uncert_estimate}
\end{align}

When $\Teff/E_{\rm eff}$ is large, the pre-factor $1/(1-(5/2)(\Teff/E_{\rm eff}))$ can be very large. This is visible in Table~\ref{tab:T0_estimates} where the relative uncertainty of $\Tmax$ can be much larger than that of $\Teff$ when $\Teff$ is extracted from photons in the lower energy ranges.

\section{Comparison with numerical calculations and implications for experimental measurements of $\Teff$}

\label{sec:comparison_with_calcs_and_data}

The mapping between $\Tmax$ and $\Teff$ derived in the previous section was validated against numerical simulations that included hydrodynamic expansion with the QCD equation of state. However, it did not include many features of modern simulations, such as fluctuating initial conditions and viscosity. Moreover, in the previous section, we focused on mapping $\Teff$ to $\Tmax$ under the assumption that $\Teff$ was extracted from a pure thermal photon spectrum. In reality, thermal photons cannot necessarily be easily isolated from non-thermal sources in data.

In this section, we use the results from state-of-the-art simulations of heavy-ion collisions to discuss the contribution of non-thermal sources to $\Teff$, and to validate the conclusions from the previous section, in particular Eq.~\ref{eq:T0_estimate}.
Using these conclusions, we then apply Eq.~\ref{eq:T0_estimate} to $\Teff$ measurements from the PHENIX Collaboration to estimate from $\Teff$ the maximum temperature of quark-gluon plasma produced in Au-Au collisions at RHIC.

\subsection{Non-thermal sources}

\label{sec:non_thermal_sources}

\paragraph{Pre-equilibrium photons}
\label{sec:profile_preeq}

In this work, we assumed that thermal photons are those photons produced during the hydrodynamic evolution of quark-gluon plasma. This is standard practice, but in reality, thermal photons do not have such a clear-cut definition. Quark-gluon plasma needs time to form after the impact of the nuclei.
Physically, the local equilibration process is likely a result of the very strong interactions among quarks and gluons, combined with the slowing expansion rate of the plasma.\footnote{The transverse expansion does steadily increase over  the lifetime of the plasma, but it never compensates for the very large decrease over time in the longitudinal expansion.}
Equilibration is unlikely to occur simultaneously across the spatially-inhomogeneous plasma, although the expansion rate at early times is strongly dominated by the longitudinal expansion which is determined by the longitudinal proper time $\tau$.
Using a fixed $\tau=\tau_0$ to divide the pre-equilibrium and hydrodynamics regions of the plasma is thus common, if imperfect. 
We have used this prescription in this work: we assumed a certain temperature profile peaked at $\Tmax$ defined at a certain time  $\tau_0$ (see Eq.~\ref{eq:T0_profile_gaussian} and text surrounding it). The energy density of the plasma certainly reaches larger values at $\tau<\tau_0$, but we can think of $\tau_0$ as the time where we are comfortable converting this energy density into a temperature under the assumption of local equilibrium.

It is natural to propagate this choice of pre-equilibrium and hydrodynamic stages to the division of pre-equilibrium photons and thermal photons. Emission of photons should be smooth across this division, in the same way that the transition between the pre-equilibrium stage and hydrodynamics should be. Additionally, chemical equilibration is relevant for photon emission. There is a gradual ramp-up of photon production in the early stage of heavy-ion collisions, as quarks are chemically equilibrating with the initial gluon-dominated medium~\cite{Traxler:1995kx,Chaudhuri:1998kv,Gelis:2004ep,Monnai:2014kqa,Bhattacharya:2015ada,Monnai:2015bca,Linnyk:2015rco,Greif:2016jeb,Vovchenko:2016ijt,Srivastava:2016hwr,Oliva:2017pri,Berges:2017eom,Monnai:2019vup,Churchill:2020uvk,Garcia-Montero:2019vju,Gale:2021emg,Garcia-Montero:2023lrd}.
It is understood that chemical and thermal equilibrium do not necessarily occur at the same time~\cite{Kurkela:2018xxd,Schlichting:2019abc,Berges:2020fwq}, which introduces additional uncertainties in defining thermal and pre-equilibrium photons. We do not attempt to resolve all these difficulties in this work, and will use a clear-cut division between pre-equilibrium and thermal photons when there is a need to understand the role of each source, with the understanding that this division is not unique. When relevant, thermal and pre-equilibrium photons will be added together, as they should be in general, removing the ambiguity in the categorization.

\paragraph{Prompt photons}
\label{sec:prompt}

Prompt photons are those produced in hard nucleon-nucleon scattering, and are the dominant source of direct photons in proton-proton collisions. They can be calculated using collinear factorization perturbative QCD as in proton-proton collisions, or by using fits extracted from direct photon measurements in proton-proton collisions. They are known to dominate the high-energy direct photon signal in nucleus-nucleus collisions, competing with thermal photons at energies $E \gtrsim 3$-$4$~GeV in $\sqrts=200$~GeV collisions or at higher LHC collision energies~\cite{Shen:2013vja,vanHees:2011vb,Chatterjee:2012dn,Paquet:2015lta,Kim:2016ylr,Dasgupta:2018pjm,Garcia-Montero:2019kjk,Monnai:2022hfs,Gale:2021emg}.
Prompt photons are expected to be different in proton-proton and nucleus-nucleus collisions. One reason is the different electric charge content of nuclei compared to protons, and difference in the nuclear parton distribution function~\cite{Arleo:2004gn,Arleo:2011gc,Helenius:2016dsk}. A second reason is the interaction of energetic partons with the quark-gluon plasma which modifies fragmentation photons and can create additional jet-medium photons; these effects are complex and are still being investigated~\cite{Zakharov:2004bi,Turbide:2005bz,Fries:2002kt,Turbide:2005fk,Turbide:2007mi,Qin:2009bk,Renk:2013kya,Yazdi:2022cuk}.

Based on the above-mentioned perturbative QCD calculations, direct photon data from proton-proton collisions and other studies, there are reasonable constraints on the prompt photon signal in nucleus-nucleus collisions, although not high-precision ones. Prompt photons are typically not subtracted from low-energy direct photon measurements in heavy-ion collisions. One recent exception is the direct photon results from the PHENIX Collaboration published in Ref.~\cite{PHENIX:2022rsx}.

For the purpose of this paper, two non-exclusive options to handle prompt photons are (i) to assume that they can be subtracted with reasonable accuracy, and (ii) focus on energy ranges where thermal photons still dominate over prompt.

\subsection{Comparison with numerical calculations}

\label{sec:comparison_with_calcs_and_expt}

\begin{figure}[t]
	\centering
	\includegraphics[width=0.99\linewidth]{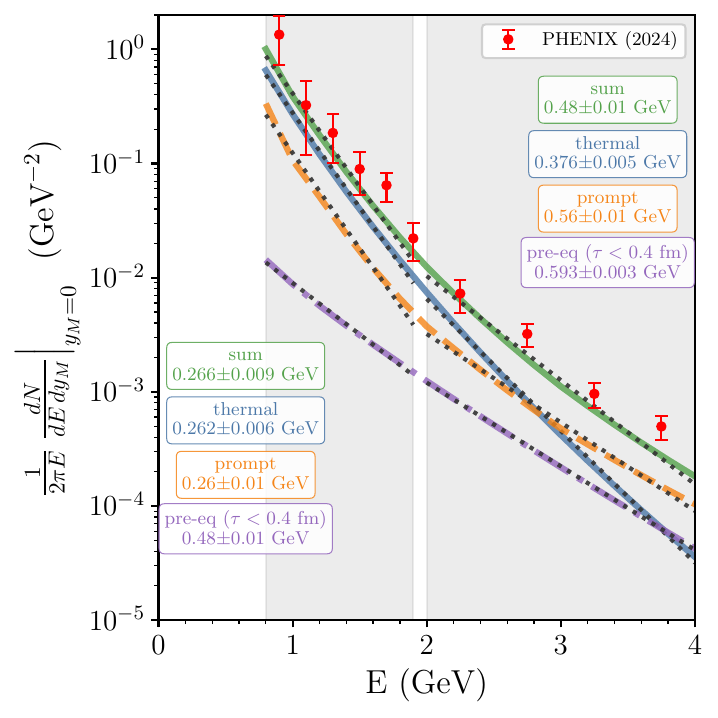}
	\caption{Prompt, pre-equilibrium and thermal photon calculations from Ref.~\cite{Gale:2021emg} compared to measurements from the PHENIX Collaboration~\cite{PHENIX:2022rsx} in Au-Au $\sqrts=200$~GeV collisions, $0$--$10$\% centrality. The greyed out areas represent the energy ranges $E \in [0.8,1.9]$~GeV and $E \in [2,4]$~GeV where the PHENIX Collaboration extracted the inverse slope. The values of the inverse slopes of the different photon sources calculated in Ref.~\cite{Gale:2021emg} are quoted to the left for $E \in [0.8,1.9]$~GeV and to the right for $E \in [2,4]$~GeV.
	}
	\label{fig:spectrum_multimessenger_phenix}
\end{figure}

Figure~\ref{fig:spectrum_multimessenger_phenix} shows direct photon measurements from PHENIX~\cite{PHENIX:2022rsx} compared with numerical calculations of prompt, thermal and pre-equilibrium photons from Ref.~\cite{Gale:2021emg}.
This is for Au-Au collisions at $\sqrts=200$~GeV with a centrality of $0$--$10$\%, which is the closest to head-on collisions for which direct photon experimental data is available for this data set. 
The statistical and systematic uncertainties of the PHENIX data are combined into one bar for simplicity. 
The calculations from Ref.~\cite{Gale:2021emg} include realistic initial conditions and pre-hydrodynamic evolution, and viscous hydrodynamics with corrections to the photon emission rates due to viscosity.\footnote{Different initial stage configurations were investigated in Ref.~\cite{Gale:2021emg}. We specifically use the case where the pre-equilibrium model K\o{}MP\o{}ST~\cite{Kurkela:2018wud,Kurkela:2018vqr} is used up to $\tau=0.4$~fm, and where photon emission is suppressed at early times to approximate a chemical equilibrium time of $\tau_{\rm chem}=1$~fm.}
The agreement of the calculations with data varies by photon energy and center-of-mass energy, but in general, they reproduce the trend but are often on the lower end of the data's uncertainty. 
Figure~\ref{fig:spectrum_multimessenger_phenix}  by itself does not necessarily reflect some of the challenges of current models in describing the centrality dependence of direct photon measurements, whether  the energy spectrum~\cite{PHENIX:2008uif,PHENIX:2009gyd,PHENIX:2014nkk,ALICE:2015xmh,STAR:2016use,PHENIX:2018for,PHENIX:2022rsx,PHENIX:2022qfp,ALICE:2023jef} or the $v_n$ momentum anisotropies~\cite{PHENIX:2011oxq,PHENIX:2015igl,PHENIX:2025ejr,ALICE:2018dti}. We set those aside, since the present paper is focused on the photon energy spectrum in central collisions.

Figure~\ref{fig:spectrum_multimessenger_phenix} also shows the inverse slopes of the different sources of photons, extracted in two different energy ranges ($E \in [0.8,1.9]$~GeV and $E \in [2,4]$~GeV) that match the ranges used by the PHENIX Collaboration for inverse slope extractions in Ref.~\cite{PHENIX:2022rsx}. Thermal photons are those discussed in Sections~\ref{sec:thermal_photons_midrap}--\ref{sec:photon_spectrum_no_doppler}, pre-equilibrium and prompt photons were discussed in Section~\ref{sec:non_thermal_sources}.

An important note is that the prompt photon calculation is shown down to an energy of $E=0.9$~GeV, but there are very good reasons to question the accuracy of the perturbative QCD calculation at such low energies. It is shown for completeness, but one should remember that the low-energy slope of prompt photons may not be reliable. The same could be said for the prompt photon calculation in the $E\in[2,4]$~GeV range, although there is at least phenomenological evidence that the perturbative QCD calculation agrees well with direct photon measurements from proton-proton collisions in that energy range~\cite{Aurenche:2006vj,PHENIX:2009gyd,PHENIX:2012jgx,Helenius:2013bya,Paquet:2015lta,ALICE:2018mjj,ALICE:2024vwy}.

As discussed in Section~\ref{sec:non_thermal_sources}, there is arbitrariness in this division between pre-equilibrium and thermal photons. In fact, studies have shown that initializing the hydrodynamics at an earlier or later time (within a reasonable range) does not change significantly the sum of pre-equilibrium and thermal photons, as long as pre-equilibrium photons are taken into account~\cite{Garcia-Montero:2023lrd,Gale:2021emg}. 
For this work, we label as pre-equilibrium photons those produced at a time $\tau<0.4$~fm, which is the hydrodynamic initialization time used for this calculation. This matches the use of the pre-equilibrium model K\o{}MP\o{}ST for $\tau<0.4$~fm in these calculations.

With Figure~\ref{fig:spectrum_multimessenger_phenix}, we attempt to answer two separate questions:
\begin{itemize}
	\item According to theoretical calculations, what is the difference between the inverse slope of thermal photons only and the inverse slope of all non-decay sources of photons (in this case: thermal, pre-equilibrium and prompt)?
	\item Assuming that we could isolate thermal photons from the total direct photon signal, do the formulas derived in Section~\ref{sec:photon_spectrum_no_doppler} under various approximations work in a more realistic setting that includes initial-state fluctuations, hydrodynamics with shear and bulk viscosity, and collisions in a finite centrality bin (here, $0$--$10$\%) rather than fully head-on?
\end{itemize}

\subsection{Effect of prompt and pre-equilibrium photons on the inverse slope $\Teff$}

\begin{table}[tb]
	\centering
	\begin{tabular}{|l|c|c|c|c|}
		\hline
		& \multicolumn{4}{|c|}{\textrm{$\Teff$}}\\
		\hline
		\shortstack{E\\range}  & Pre-eq $\Teff$ & Thermal & \shortstack{Thermal\\+Pre-eq} & \shortstack{Thermal\\+Pre-eq\\+Prompt} \\
		\textrm{[GeV]} & [GeV] & [GeV] & [GeV] & [GeV] \\
		\hline
		0.8–1.9 & $0.48\pm0.01$ & $0.262\pm0.006$ & $0.269\pm0.007$ & $0.266\pm0.009$ \\
		\hline
		2.0–4.0 & $0.593\pm0.003$ & $0.376\pm0.005$ & $0.427\pm0.009$ & $0.48\pm0.01$ \\
		\hline
		\hline
		0.8--4.0 & $0.559\pm0.007$ & $0.332\pm0.008$ & $0.36\pm0.01$ & $0.38\pm0.01$ \\
		\hline
	\end{tabular}
	\caption{Inverse slope $\Teff$ in GeV extracted from Au-Au $\sqrts=200$~GeV $0$--$10$\% centrality calculations from  Ref.~\cite{Gale:2021emg} over different energy ranges for thermal photons, pre-equilibrium photons, and their sum with and without prompt photons. 
	The uncertainty is the standard error from the linear regression. 	
		See text and Figure~\ref{fig:spectrum_multimessenger_phenix} for details.
		}
	\label{tab:Teff_AuAu200_cent0010_mutimessenger}
\end{table}

Table~\ref{tab:Teff_AuAu200_cent0010_mutimessenger} lists the inverse slope $\Teff$ for different combinations of sources of photons shown in Figure~\ref{fig:spectrum_multimessenger_phenix}. The pre-equilibrium and thermal photon slopes are the same as those shown in the figure. The other two slopes are those extracted from (i) the sum of thermal and pre-equilibrium photons, and (ii) the sum of thermal, pre-equilibrium and prompt photons. 
We focus on these two additional inverse slopes, since they provide a quantification of the impact of  pre-equilibrium and prompt photons on the inverse slope.
The inverse slopes are quoted for the same energy ranges as shown in Figure~\ref{fig:spectrum_multimessenger_phenix} ($E \in [0.8,1.9]$~GeV and $E \in [2,4]$~GeV), as well as for the whole range ($E \in [0.8,4]$~GeV).

The lower energy range, $E \in [0.8,1.9]$~GeV, is the one with the least competition between thermal photons and pre-equilibrium photons. The contribution of prompt photons is less certain, because prompt photon calculations are much less reliable there, and there is little data from proton-proton collisions to constrain it at the moment. According to current estimates of prompt photons, thermal photons are nevertheless the dominant contribution in this energy range.
The effect of pre-equilibrium photons and prompt photons on $\Teff$ is small, changing from $0.262$~GeV for only thermal photons, to $0.269$~GeV with the addition of pre-equilibrium photons, to $0.266$~GeV with the additional inclusion of prompt photons, an overall effect below $3$\%.  

In the higher photon energy range, $E \in [2,4]$~GeV, the effect of pre-equilibrium and prompt photons is much larger, as can be expected. The inverse slope of thermal photons $\Teff=0.376$~GeV increases by $14$\% to $0.427$~GeV with the inclusion of pre-equilibrium photons, and then the inclusion of prompt photons increases it by another $12$\% to $0.48$~GeV. While these factors may not seem large, if we insert these values in Eq.~\ref{eq:T0_estimate}, this is the difference between an estimated maximum temperature of $0.558$~GeV and $0.822$~GeV. A similar conclusion can be drawn from Eq.~\ref{eq:T0_uncert_estimate}: a relative $30$\% uncertainty on $\Teff$ translates into a relative uncertainty of at least $50$\% on $\Tmax$, simply because the photon energy is not much larger than $\Teff$.

This would point to using lower-energy photons where $\Teff$ has less contamination from non-thermal sources. Unfortunately, even under simplified conditions, we saw in Section~\ref{sec:photon_spectrum_no_doppler} that it is difficult to relate $\Teff$ measured at low $E$ to the maximum temperature of the plasma.

\subsection{Comparing the inverse slope $\Teff$ to the plasma's maximum temperature}

In Section~\ref{sec:photon_spectrum_no_doppler}, we concluded that the plasma's maximum temperature could be estimated well from the inverse slope of the thermal photon energy spectrum, as long as the slope is extracted from sufficiently high-energy photons. Setting aside any possible contamination from pre-equilibrium photons, and subtleties related to prompt photons, we verify numerically here whether the mapping between $\Tmax$ and $\Teff$ from Section~\ref{sec:photon_spectrum_no_doppler} still holds when a realistic simulation of quark-gluon plasma is performed. As a reminder,  Section~\ref{sec:photon_spectrum_no_doppler} neglected the effect of viscosity, of initial-state fluctuations, or of centrality dependence, while the calculation from Ref.~\cite{Gale:2021emg} includes all these effects.

Because we now look at calculations with fluctuating initial conditions, a single $\Tmax$ is not defined. We can nevertheless compute an averaged maximum temperature from simulations, which we obtain by finding the maximum temperature of each event and averaging these maximum temperatures. This is the maximum temperature at the $\tau=0.4$~fm initialization time of hydrodynamics. The value from simulations is approximately $\Tmax \approx 0.54$~GeV.

\begin{table}[tb]
	\centering
	\begin{tabular}{|l|c|c|c|}
		\hline
		$E$ range & $E_{\rm eff}$  & thermal $\Teff$ & $\Tmax$ estimate\\
		(GeV) & (GeV) & (GeV) & (GeV) \\
		\hline
		0.8–1.9 & 1.27 & $0.262\pm0.006$ & $0.54 \pm 0.03$ \\
		\hline
		2.0–4.0 & 2.89 & $0.376\pm0.005$ & $0.56\pm 0.01$ \\
		\hline
		\hline
		0.8–4.0 & 1.99  & $0.332\pm0.008$ & $0.57 \pm  0.02$ \\
		\hline
	\end{tabular}
	\caption{Maximum temperature $\Tmax$ estimated from the inverse slope $\Teff$ extracted over different energy ranges for thermal photons only. 
	The uncertainty is the standard error from the linear regression; it is \emph{not} a theoretical uncertainty. 
		The effective energy $E_{\rm eff}$ of the photon in the energy range, Eq.~\ref{eq:Eeff}, is also quoted.}
	\label{tab:T0_from_Teff_AuAu200_cent0010_mutimessenger}
\end{table}

Applying Eq.~\ref{eq:T0_estimate} to the inverse slopes $\Teff$ extracted from  \emph{only}  thermal photons leads to the results shown in Table~\ref{tab:T0_from_Teff_AuAu200_cent0010_mutimessenger}. We see that the temperature estimates are consistent with each other, between $0.54$ and $0.57$~GeV, which is close to the maximum temperature of the plasma known from the numerical simulations. While this is only one example of numerical simulations, it suggests that Eq.~\ref{eq:T0_estimate} could be useful to estimate the maximum temperature of the plasma, and is  more reliable and interpretable as an actual temperature than $\Teff$ itself.

Our simple formula (Eq.~\ref{eq:T0_estimate}) predicts that $\Teff[E_1,E_2]$ measured in different energy ranges are related to each other by
\begin{align}
	\frac{\Teff[E_3,E_4]}{\Teff[E_1,E_2]} = 1-\frac{5}{2} \Teff[E_3,E_4] \left(  \frac{\ln\left(\frac{E_4}{E_3}\right)}{E_4-E_3}-\frac{\ln\left(\frac{E_2}{E_1}\right)}{E_2-E_1}  \right)
	\label{eq:Teff_E_dep}
\end{align}

This can be compared with the calculations from Ref.~\cite{Gale:2021emg} (Table~\ref{tab:Teff_AuAu200_cent0010_mutimessenger}). In that case, the thermal photon $\Teff$ increases by $40$\% between $E \in [2,4]$~GeV and $[0.8,1.9]$~GeV.
Equation~\ref{eq:Teff_E_dep} predicts that the inverse slope should increase by a factor of $1.41$, which agrees almost perfectly.
These results suggest that the analytical formulas derived in this work (Eq.~\ref{eq:Teff_E_dep} and the various results from Section~\ref{sec:photon_spectrum_no_doppler}) provide the right trends to understand the calculated and measured $\Teff$.

\subsection{Discussion of effective temperature extractions from collider measurements}

The PHENIX Collaboration published  in Ref.~\cite{PHENIX:2022rsx} photon spectra measurements with prompt photons subtracted from direct photons, which they refer to as ``nonprompt direct photons''. In what follows, we use these measurements to study the inverse slope $\Teff$ in Au-Au in $\sqrts=200$~GeV collisions with $0$--$10$\% centrality, the most central collisions available for this dataset.

The inverse slope measured by the PHENIX Collaboration for $0$--$10$\% centrality for $E \in [0.8,1.9]$~GeV is $0.27 \phantom{ }^{+0.04}_{-0.03}$~GeV, which is very similar to the values extracted from the theoretical calculations of Ref.~\cite{Gale:2021emg} quoted in Table~\ref{tab:Teff_AuAu200_cent0010_mutimessenger}. As discussed earlier in this section, the calculations return a similar inverse slope whether it is calculated with only thermal photons, or including pre-equilibrium and prompt photons.

For $E \in [2,4]$~GeV, the inverse slope measured by PHENIX is $\Teff=0.51  \phantom{ }^{+0.09}_{-0.07}$~GeV. The  $\Teff$ from hydrodynamic calculations quoted in Table~\ref{tab:Teff_AuAu200_cent0010_mutimessenger} range from $0.376$~GeV for only thermal photons  to $0.427$~GeV if pre-equilibrium photons are included. Thus, the calculated $\Teff$ is outside of the error bars of the PHENIX measurements. 
Differences between theoretical calculations of $\Teff$ and PHENIX's measurements are not unexpected, since in general, there is known tension between calculations and the measured spectrum and momentum anisotropies.
What the current analysis adds to the discussion is that pre-equilibrium photons have a significant impact on $\Teff$, raising $\Teff$ by approximately $0.05$~GeV in this case.

The PHENIX nonprompt $\Teff$  in $E \in [2,4]$~GeV is almost double that in $E \in [0.8,1.9]$~GeV ($0.51$ vs $0.27$~GeV). 
The state-of-the-art calculations from Ref.~\cite{Gale:2021emg} predict a 60\% increase of $\Teff$ extracted in these two bins. This increase is actually 40\% for thermal photons only, but 60\% once thermal and pre-equilibrium photons are added, which should be the right value to compare with measurements. 

Given that the calculations were already known to have tension with the PHENIX measurements, this observation is not necessarily surprising. On the other hand, tension between calculations and data is often quantified in terms of the ratio of the respective spectrum; $\Teff$ highlights the differences in the slope rather than normalization, which is useful additional information. 

When it comes to $\Teff$ extracted from measurements in general, most measurements do not subtract prompt photons when evaluating $\Teff$. For $E \in [2,4]$~GeV, according to calculations from Ref.~\cite{Gale:2021emg}, including non-thermal sources increases $\Teff$ by a full $0.1$~GeV, and as such should not be directly associated with a QGP temperature. Because pre-equilibrium and prompt contribute much less at low photon energies, there is a risk of attributing differences across photon energy range to temperature effects while they are partly due to non-thermal effects.
Importantly, prompt photons are very likely modified in heavy-ion collisions~\cite{Zakharov:2004bi,Turbide:2005bz,Fries:2002kt,Turbide:2005fk,Turbide:2007mi,Qin:2009bk,Renk:2013kya,Yazdi:2022cuk}, and subtracting prompt photon estimates from measurements can also bias the extracted $\Teff$.

\subsection{Estimating the maximum QGP temperature from collider measurements}

As discussed above, extracting the maximum temperature of the plasma from $\Teff$ measurements presents multiple challenges. If $\Teff$ is extracted at high $E$, it is very likely not a measurement of the thermal photon $\Teff$, since there are contributions from pre-equilibrium photons and prompt photons. Even if prompt photons are subtracted, it is possible this subtraction is imperfect, because QGP effects on prompt photons are still not known accurately.
If $\Teff$ is extracted at low $E$, the simple mapping between the thermal photon's $\Teff$ and $\Tmax$ does not work as well, as discussed in Section~\ref{sec:photon_spectrum_no_doppler}.
We established the approximate criterion $E_\textrm{eff}/\Teff \gtrsim 5$ in Eq.~\ref{eq:Eeff_estimate} for the range of validity of our $\Tmax \leftrightarrow \Teff$ mapping.

From the PHENIX nonprompt measurements~\cite{PHENIX:2022rsx}, for $E \in [0.8,1.9]$~GeV, we have $E_\textrm{eff}/\Teff\approx 1.27/0.27 \approx 4.7$, which is at the edge of applicability according to Eq.~\ref{eq:Eeff_estimate}. Applying Eq.~\ref{eq:T0_estimate} to $\Teff=0.27 \phantom{ }^{+0.04}_{-0.03}$~GeV, we find $\Tmax=0.58 \phantom{ }^{+0.22}_{-0.12}$~GeV.

\begin{figure}
	\centering
	\includegraphics[width=0.99\linewidth]{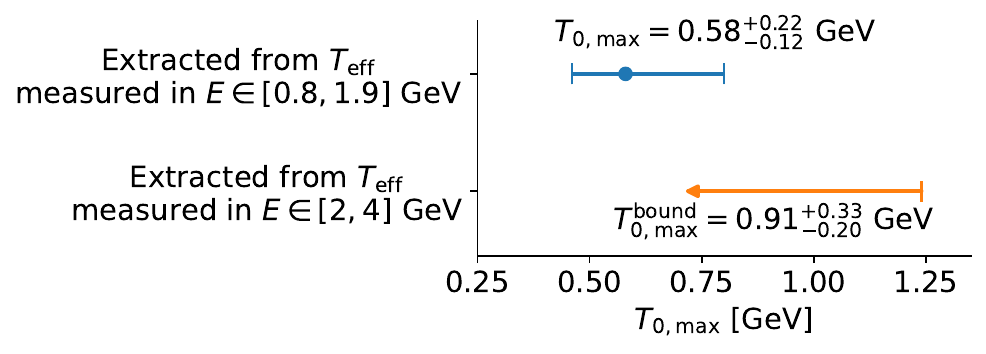}
	\caption{
	Estimate of the maximum temperature of the quark-gluon plasma in Au-Au collisions at $\sqrts=200$~GeV, $0$--$10$\% centrality, obtained using Eq.~\ref{eq:T0_estimate} with PHENIX $\Teff$ measurements~\cite{PHENIX:2022rsx} for nonprompt photons. See text for details.
	}
	\label{fig:T0max_phenix_AuAu0010}
\end{figure}

For the $\Teff$ measurement in   $E \in [2,4]$~GeV, to use Eq.~\ref{eq:T0_estimate}, we either have to (i) assume that $\Teff$ is purely thermal and extract $\Tmax$, or (ii) use Eq.~\ref{eq:T0_estimate}  with the understanding that the resulting $\Tmax$ is likely a significant overestimate.
The second case makes the assumption that pre-equilibrium and prompt photons increase $\Teff$, which should be reasonable assumptions (see first part of this section). Using Eq.~\ref{eq:T0_estimate} with $\Teff=0.51  \phantom{ }^{+0.09}_{-0.07}$~GeV, we find $\Tmax^{\textrm{bound}}=0.91 \phantom{ }^{+0.33}_{-0.20}$~GeV.

The results are displayed in Figure~\ref{fig:T0max_phenix_AuAu0010}.
As discussed in Section~\ref{sec:photon_spectrum_no_doppler} and Eq.~\ref{eq:T0_uncert_estimate}, the relative uncertainties on $\Tmax$ are considerably larger than on $\Teff$, limiting the precision at which the plasma temperature can be inferred.\footnote{While we do not add any theoretical uncertainty from the imperfect accuracy of Eq.~\ref{eq:T0_estimate}, one could argue for its inclusion. From Tables~\ref{tab:T0_estimates} and \ref{tab:T0_from_Teff_AuAu200_cent0010_mutimessenger}, an additional $10$ to $20$\% theoretical uncertainty on $\Tmax$ could be reasonable.}
Nevertheless, $\Tmax=0.58 \phantom{ }^{+0.22}_{-0.12}$~GeV extracted from $\Teff$ measured in $E \in [0.8,1.9]$~GeV is a reasonable result consistent with the maximum temperatures identified from hydrodynamic simulations. On the other hand, the estimated upper bound of $\Tmax^{\textrm{bound}}=0.91 \phantom{ }^{+0.33}_{-0.20}$~GeV is very large and provides only a very weak bound on the plasma's temperature. This is a clear challenge of interpreting a $\Teff$ that may have significant contributions from non-thermal sources.

The results from Figure~\ref{fig:T0max_phenix_AuAu0010} are broadly consistent with temperature profiles constrained from soft hadron measurements, such as in Ref.~\cite{Gale:2021emg}. While the constraints from Figure~\ref{fig:T0max_phenix_AuAu0010} have large uncertainties, we point out that they are experimentally orthogonal to those obtained from hadronic observables, being purely from electromagnetic origin. They are, however, not purely independent of numerical simulations of photon production in relativistic heavy-ion collisions, since the latter needed to be used to understand the contribution of non-thermal sources, as well as validate the range of applicability of Eq.~\ref{eq:T0_estimate}.

\section{Centrality and center-of-mass energy dependence of the photon spectrum}

\label{sec:centrality_vs_sqrts}

\begin{figure}
	\centering
	\includegraphics[width=0.72\linewidth]{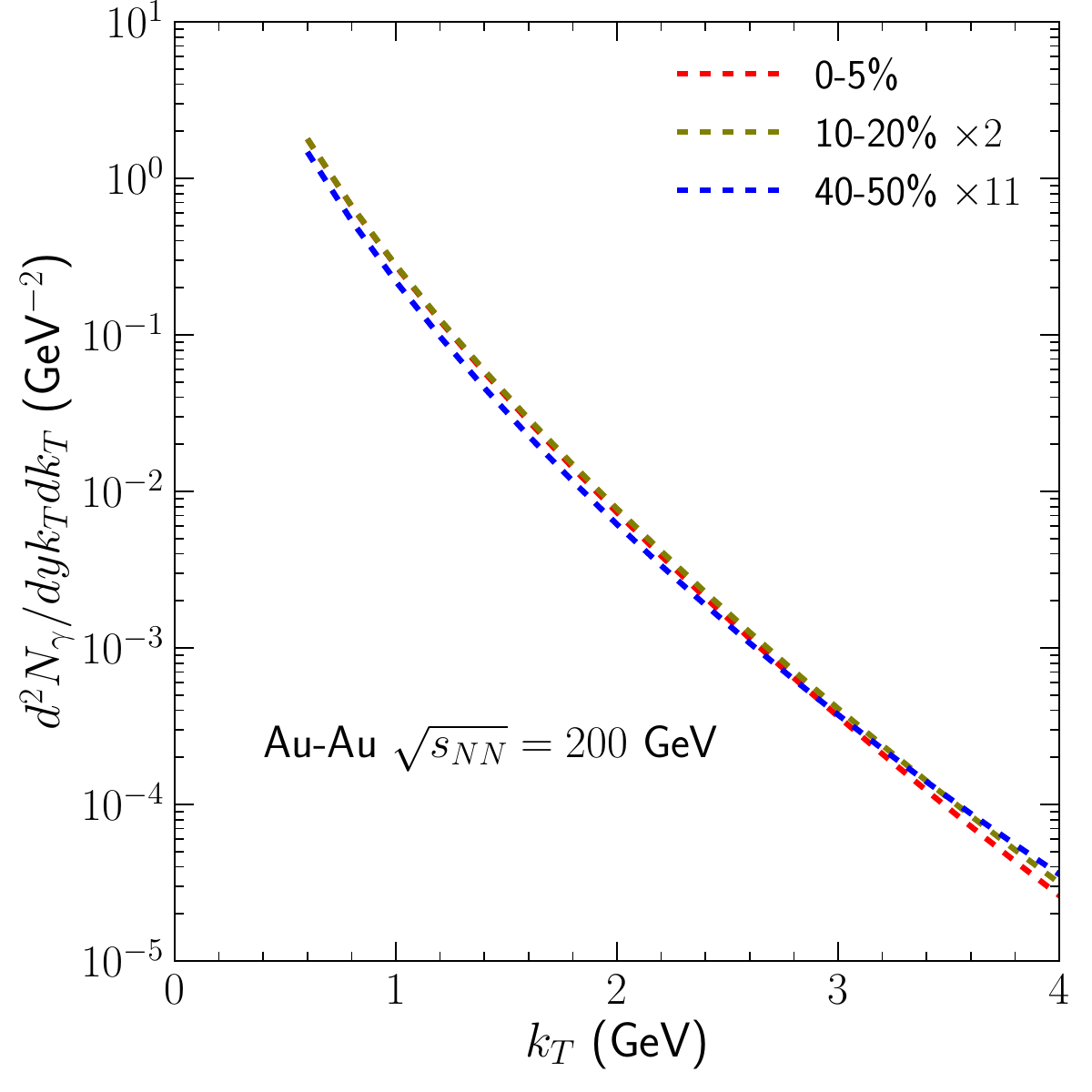}
	\includegraphics[width=0.72\linewidth]{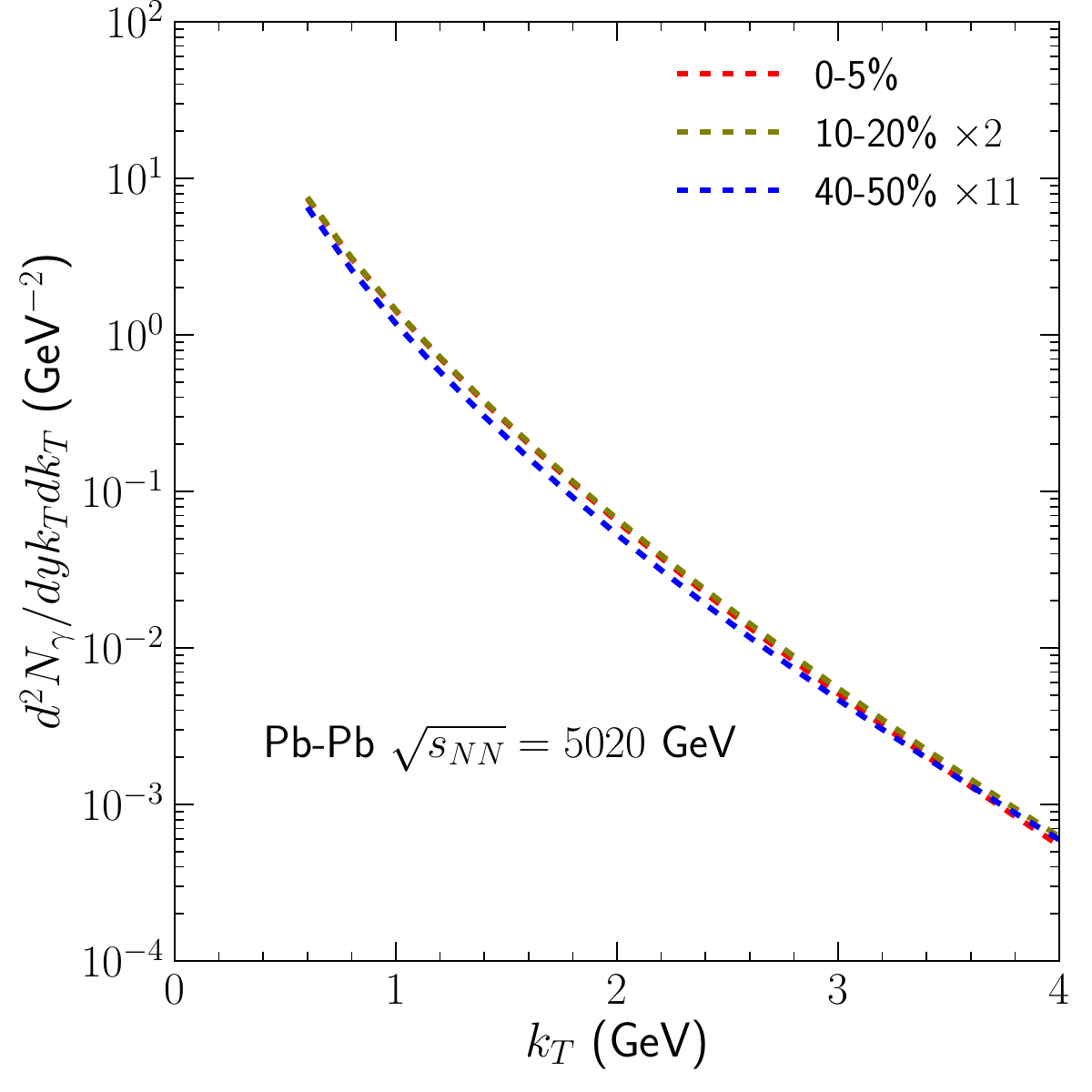}
	\includegraphics[width=0.72\linewidth]{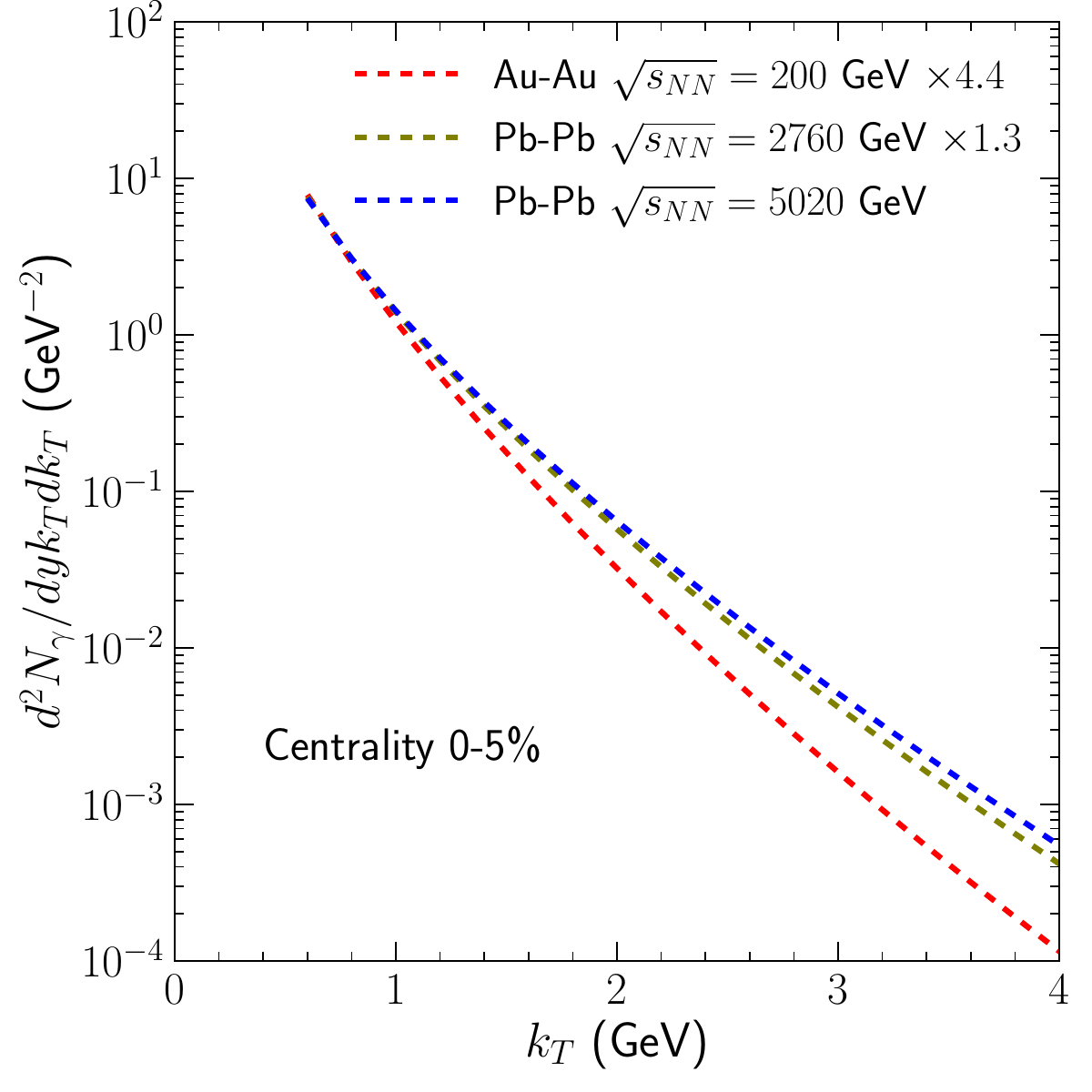}
	\caption{
		Thermal photon spectrum as a function of $k_T$ (which is equal to the photon energy $E$ at midrapidity) from Ref.~\cite{Gale:2021emg}, normalized such that they have approximately the same magnitude in the lowest $k_T$ bin. The top two panels show three different centralities ($0$--$5$\%, $10$--$20$\%, $40$--$50$\%), for Au-Au collisions at $\sqrts=200$~GeV (top) and Pb-Pb collisions at $\sqrts=5020$~GeV (middle). These two panels show that thermal photons produced in different centralities have approximately the same $k_T$ dependence, despite having very different normalizations. The bottom panel compares the 0-5\% centrality for three different collisions: Au-Au collisions at $\sqrts=200$~GeV, and Pb-Pb collisions at $\sqrts=2760$ and $5020$~GeV. In this case, the thermal photon spectrum has both a different normalization and $k_T$ dependence, as assumed in this work.
	}
	\label{fig:sqrts_vs_cent_thermal_photon_spectrum}
\end{figure}

The present work focused on head-on relativistic heavy-ion collisions at different center-of-mass collision energies. This choice was made because the different assumptions made in this work are expected to work best in collisions that are as central as possible (smallest impact parameter), for reasons discussed in the introduction. Because of this choice, we did not attempt to study the centrality dependence of the photon spectrum. 
This is a significant limitation, since the centrality dependence of the photon spectrum is an important observable to understand. However, we show in this section that numerical simulations of heavy-ion collisions indicate that the centrality dependence of the photon spectrum is  different from its center-of-mass energy dependence.

Figure~\ref{fig:sqrts_vs_cent_thermal_photon_spectrum} shows the calculations from the state-of-the-art numerical simulations of Ref.~\cite{Gale:2021emg}. The top panel shows the thermal photon spectrum for Au-Au collisions at $\sqrts=200$~GeV for $0$--$5$\%, $10$--$20$\% and $40$--$50$\% centrality, but with the last two centralities rescaled respectively by a factor of $2$ and $11$ such that their lowest $k_T$ bin is approximately the same. An equivalent plot is shown in the middle panel for Pb-Pb collisions at $\sqrts=5020$~GeV. These two plots show that the spectrum varies primarily in normalization as a function of centrality, without significant differences in the slope. This was also observed in Ref.~\cite{Massen:2024pnj}, and to some extent in Ref.~\cite{Shen:2013vja}, for example. On the other hand, the lower panel shows the photon spectrum in $0$--$5$\% centrality as a function of the center-of-mass collision energy of three systems: Au-Au at $200$~GeV, Pb-Pb at $2760$~GeV and Pb-Pb at $5020$~GeV; the first two systems are normalized by an overall factor of $4.4$ and $1.3$ respectively, such that the lower energy range of the three spectra has the same normalization. In this case, one can clearly see the change of slope, consistent with the present work's discussion that the slope reflects the initial temperature of the system.

Figure~\ref{fig:sqrts_vs_cent_thermal_photon_spectrum} shows that discussions of the interpretation of the slope of the photon spectrum should be differentiated between the center-of-mass energy dependence, and the centrality dependence.

\section{Summary}

We studied the inverse slope of the photon energy spectrum with a simplified model of quark-gluon plasma expansion with exact azimuthal symmetry and a smooth initial temperature profile, approximations that work best in central heavy-ion collisions. This simpler model retains major features affecting the inverse slope of the photon spectrum, including Doppler shift from transverse expansion, and an evolving temperature profile in space and time. We differentiated carefully between the local and global effects of the transverse Doppler shift, emphasizing that the inverse slope is only sensitive to the global effect. In particular, we showed that the Doppler shift tends to increase the inverse slope of the thermal photon spectrum at low photon energy, and decrease it at higher photon energies. This is contrary to the common assumption that Doppler shift systematically increases the inverse slope, but it is a natural consequence of two competing effects (Section~\ref{sec:effect_of_transverse_flow}): (i) Doppler shift is not significant for very low-energy photons and (ii) high-energy thermal photons tend to be emitted at early times when the transverse Doppler shift is small. 
We showed that the overall magnitude of the Doppler shift on $\Teff$ ranged from $-10$\% to $+20$\%, often much smaller.
We found that these conclusions remained valid in more complete calculations of thermal photons (Appendix~\ref{sec:Teff_prc}).
Our results show that it is crucial to differentiate between the effect of the transverse Doppler shift on the spectrum and on the inverse slope: as shown in Figure~\ref{fig:spectrum_with_without_flow}, counter-intuitively, the photon energy where the spectrum is most affected by the Doppler shift is also the point where the Doppler shift has no effect on the inverse slope.

Given the moderate effect of the Doppler shift on the inverse slope, we further simplified the problem by neglecting the Doppler effect and deriving an expression relating the inverse slope of \emph{thermal} photons to the maximum temperature of the quark-gluon plasma (Eq.~\ref{eq:T0_estimate}).
We quantified non-exponential corrections to the photon energy spectrum that are independent of the transverse Doppler shift (Eq.~\ref{eq:slope}).

We validated the equations derived under simplifying assumptions by comparing to state-of-the-art thermal photon calculations (Section~\ref{sec:comparison_with_calcs_and_expt}). We found that Eq.~\ref{eq:T0_estimate}, and the intuition from Section~\ref{sec:photon_spectrum_no_doppler}, were consistent with the results of numerical calculations of thermal photons that include the effect of viscosity, realistic heavy-ion initial conditions, and more sophisticated thermal photon emission rates. Using the same numerical calculations, we studied the effect of pre-equilibrium photons and prompt photons on the inverse slope, and found that contamination from these non-thermal photon sources can exceed effects such as transverse Doppler shifts.

Our work highlights both the potential and the limits of attempting to map $\Teff$ to the plasma's temperature. From Section~\ref{sec:photon_spectrum_no_doppler} (Table~\ref{tab:T0_estimates}) and Section~\ref{sec:comparison_with_calcs_and_data} (Table~\ref{tab:T0_from_Teff_AuAu200_cent0010_mutimessenger}), we found that we could map $\Teff$ to the plasma's maximum temperature with relatively good accuracy, at least as long as the inverse slope is measured at sufficiently high photon energies (Eq.~\ref{eq:Eeff_estimate}).
We believe this is an important result: $\Teff$ being the result of a superposition of photons emitted at different temperatures and with different Doppler shifts does not preclude mapping it to $\Tmax$, at least in central collisions.
The contribution of non-thermal photon sources to $\Teff$ appears to be a more significant challenge. We emphasized this in our analysis of PHENIX $\Teff$ measurements~\cite{PHENIX:2022rsx}, where we estimated $\Tmax=0.58 \phantom{ }^{+0.22}_{-0.12}$~GeV from the $E \in [0.8,1.9]$~GeV $\Teff$ measurement, but only extracted an upper estimate ($\Tmax^{\textrm{bound}}=0.91 \phantom{ }^{+0.33}_{-0.20}$~GeV) from the $E \in [2,4]$~GeV $\Teff$
(Figure~\ref{fig:T0max_phenix_AuAu0010}). Pre-equilibrium photon production is deeply intertwined with the very definition of a maximum temperature, since photons are sensitive to the complex question of the local thermal and chemical equilibration of the plasma.
Trying to avoid non-thermal sources by focusing on lower energy photons appears a straightforward solution, but we found that even in our simplified model, we could not reliably map $\Teff$ measured at lower photon energies ($E \lesssim 5 \Teff$) to $\Tmax$. 
Moreover, the low-energy spectrum is not immune to non-thermal sources: it receives contributions from the late, dilute hadronic stage, where non-equilibrium effects can be significant~\cite{Huovinen:2002im,Baeuchle:2009ep,Gotz:2021dco}.

The present work aims to complement insights from state-of-the-art numerical simulations, which can model fluctuating initial conditions, hydrodynamics with both shear and bulk viscosity and corresponding corrections to the photon emission rates, medium effects on fragmentation photons and other photon production mechanisms. Modern simulations can provide model-dependent numerical mappings between $\Teff$ and various averages of the fluctuating hydrodynamic temperature profiles $T(\tau,x,y,\eta_s)$. The present work represents a limiting case for a  $\Teff \leftrightarrow \Tmax$ mapping, with more limited applicability (focus on higher photon energies and central collisions) but with simple and transparent assumptions. Moreover, the simplified model allowed us to better isolate effects that are present in state-of-the-art simulations. For example, the counter-intuitive effect on $\Teff$ of the transverse Doppler shift appears consistent with the results of more sophisticated calculations (Appendix~\ref{sec:Teff_prc}), but our model makes it especially straightforward to identify.

An important future direction would be an improved treatment of pre-equilibrium photons~\cite{Chaudhuri:2011up,Bhattacharya:2015ada,Vovchenko:2016ijt,Srivastava:2016hwr,Berges:2017eom,Monnai:2019vup,Kasmaei:2019ofu,Garcia-Montero:2019vju,Churchill:2020uvk,Almaalol:2020rnu,Gale:2021emg,Garcia-Montero:2023lrd} and medium effects on prompt photons~\cite{Shen:2013vja,vanHees:2011vb,Chatterjee:2012dn,Paquet:2015lta,Kim:2016ylr,Dasgupta:2018pjm,Garcia-Montero:2019kjk,Monnai:2022hfs,Gale:2021emg}, which are essential if one wants to interpret the inverse slope of the photon spectrum extracted for $E \gtrsim 3$--$4$~GeV.
A more challenging but important generalization of this work would study the centrality dependence of the inverse slope. %
Finally, applying our approach to virtual photons (lepton pairs) is a natural extension.

\acknowledgments
This work was supported by the U.S. Department of Energy, Office of Science under Award Numbers DE-SC-0024347 and DE-FG02-05ER41367. 
We thank Mike Sas, Marco van Leeuwen and the ALICE electromagnetic physics working group for questions and discussions that led to this manuscript.
We thank Axel Drees, Charles Gale, Wenqing Fan, Matthew Heffernan, Ulrich Heinz, Vladimir Khachatryan, Dananjaya Liyanage, Guy Moore, Berndt M\"uller, Anthony Muñoz, Klaus Reygers, Bj\"orn Schenke, Chun Shen and Dinesh Srivastava for discussions and valuable feedback on the manuscript. 
This research used resources of the National Energy Research Scientific Computing Center (NERSC), a U.S. Department of Energy Office of Science User Facility operated under Contract No. DE-AC02-05CH11231. We gratefully acknowledge technical support from Rollin Thomas at NERSC.
ChatGPT 5 and Claude Opus 4 were used to write and improve some plotting and analysis scripts, and to proofread the manuscript.

\appendix

\section{Deviations of the thermal photon rate from a pure exponential}

\label{sec:appendix_rate_exp}

\begin{figure}[tb]
	\centering
	\includegraphics[width=\linewidth]{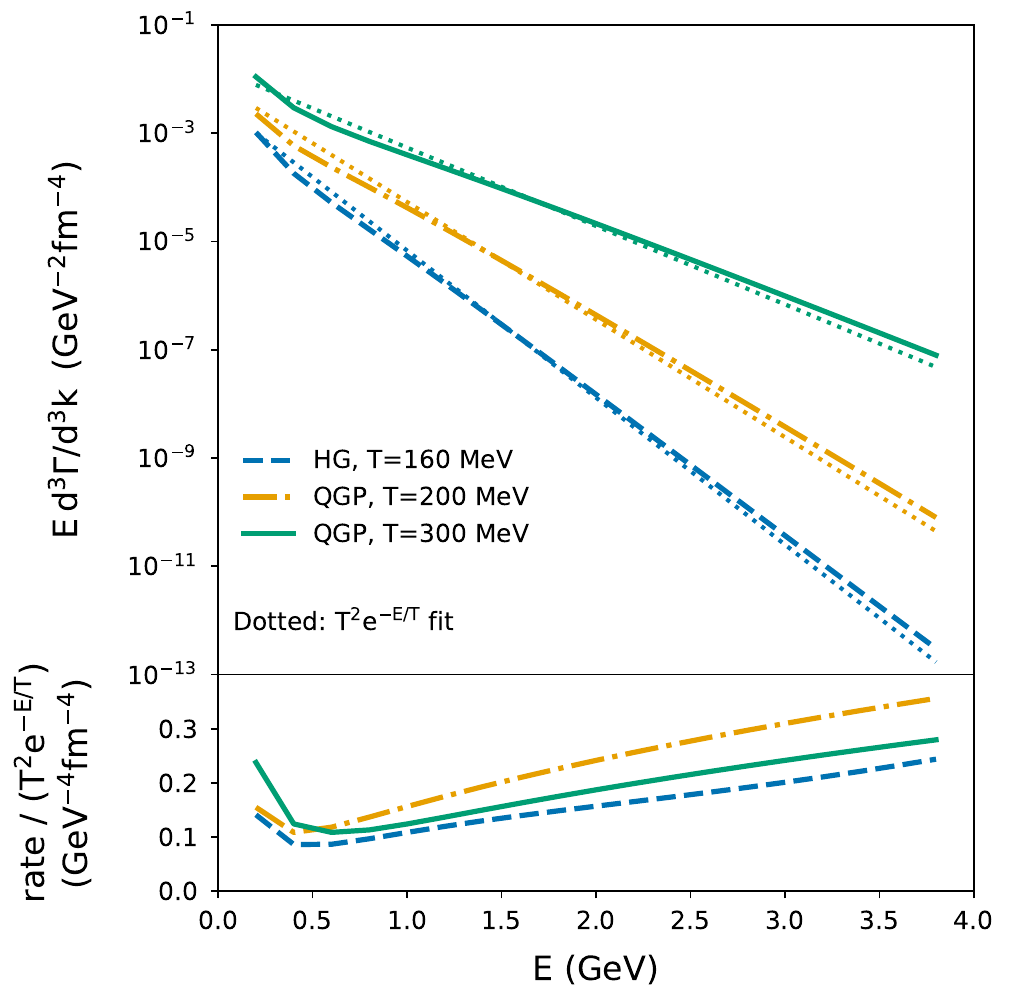}
	\caption{Top panel: Thermal photon emission rate for a leading-order weakly-coupled QGP~\cite{Arnold:2001ms} ($T=200, 300$~MeV) or for a hadronic gas~\cite{Turbide:2003si,Liu:2006imd,Heffernan:2014mla,Holt:2015cda} ($T=160$~MeV) compared with $T^2 e^{-E/T}$ (normalized to have the right magnitude), as a function of the photon energy $E$.  Bottom panel: ratio of the thermal rate to  $T^2 e^{-E/T}$.
	}
	\label{fig:rate_vs_expt}
\end{figure}

The photon emission rate of thermalized nuclear matter at zero baryon density has been computed in multiple different limits, including for a high-temperature weakly-coupled gas of quarks and gluons~\cite{Kapusta:1991qp, Aurenche:1998nw, Arnold:2001ms,Ghiglieri:2013gia,Gale:2014dfa,Arnold:2001ms,Ghiglieri:2013gia,Hidaka:2015ima} and for an interacting system of hadrons~\cite{Kapusta:1991qp, Turbide:2003si,Liu:2006imd,Dusling:2009ej,Lee:2014pwa,Holt:2015cda,Holt:2020mwf}. In this work, we use the fact that the dominant feature of the energy $E$ and temperature $T$ dependence of the rate is
\begin{align*}
k \frac{d^3 \Gamma_\gamma(E,T)}{d^3 k}\sim T^2 e^{-E/T}
\end{align*}
with the remaining $E$ and $T$ dependence being minor factors \emph{compared} to these. 

To quantify the quality of this approximation, we show in Figure~\ref{fig:rate_vs_expt} the rate compared to $T^2 e^{-E/T}$ at three temperatures, $T=160$, $200$ and $300$~MeV. For the two higher temperatures, the rate is given by the leading-order weakly-coupled QGP calculations from Ref.~\cite{Arnold:2001ms}, while for the lower temperature, the rate is given by the hadronic calculations from Refs.~\cite{Turbide:2003si,Liu:2006imd,Heffernan:2014mla,Holt:2015cda}.

Figure~\ref{fig:rate_vs_expt} shows that the exponential factor captures almost the entire variation of the rate in energy, and what is left is a slowly varying factor that varies by a factor of approximately $2$ to $3$ while the rate varies by multiple orders of magnitude.

\section{Speed of sound, transverse expansion, and the unexpected accuracy of conformal Bjorken hydrodynamics}

\label{sec:appendix_dVdT}

The expansion of quark-gluon plasma produced in heavy-ion collisions is initially dominated by longitudinal expansion along the collision axis. Transverse expansion builds up with time as a result of interactions. As transverse expansion increases and longitudinal expansion decreases, the system transitions from a nearly one-dimensional initial expansion to a three-dimensional expansion at later times.

During the initial one-dimensional expansion, the plasma is still relatively hot, and its speed of sound should not be far from QCD's conformal limit of $c_s^2=1/3$. The solution of inviscid relativistic hydrodynamics in the regime of constant speed of sound and one-dimensional longitudinal expansion is the well-known Bjorken solution~\cite{Bjorken:1982qr}:
\begin{align}
T(\tau)=\Tmax (\tau_0/\tau)^{c_s^2}.
\end{align}
It is a good solution as long as the speed of sound does not vary significantly, or the longitudinal expansion dominates over the transverse one. For the plasma produced in nuclear collisions, both approximations are broken at later times: the speed of sound of QCD is only approximately constant at high temperature (above $\approx 0.4$--$0.5$~GeV), while the transverse flow builds up with time.

\begin{figure}
	\centering
	\includegraphics[width=0.72\linewidth]{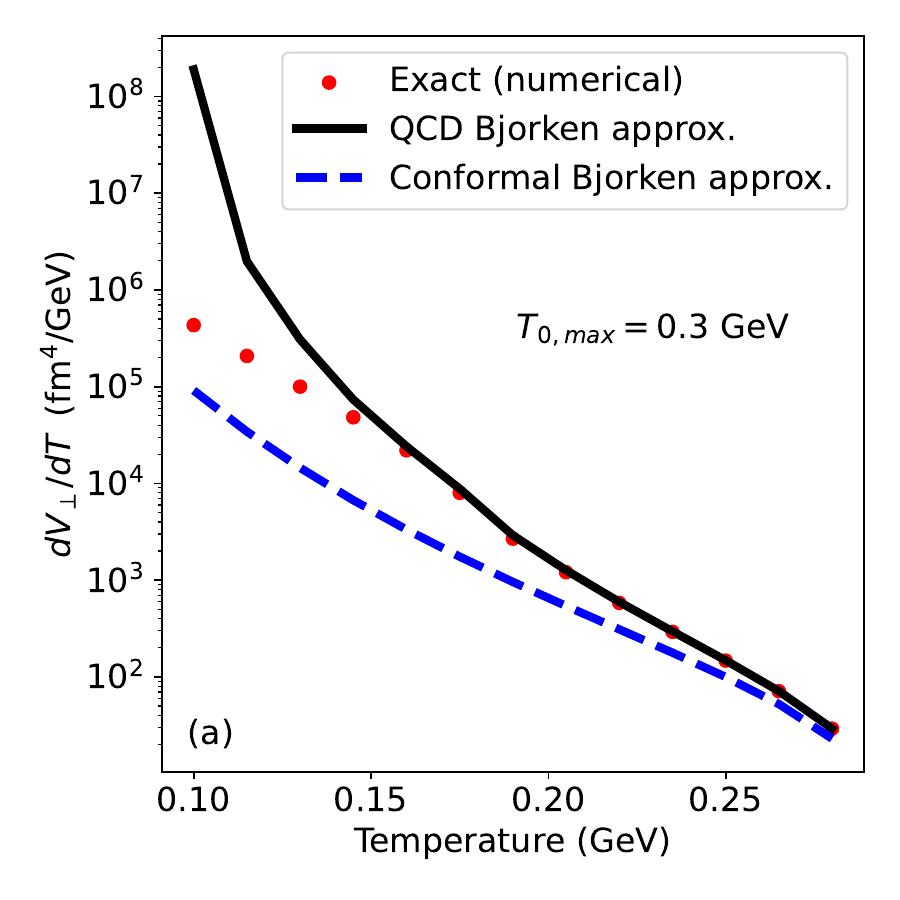}
	\includegraphics[width=0.72\linewidth]{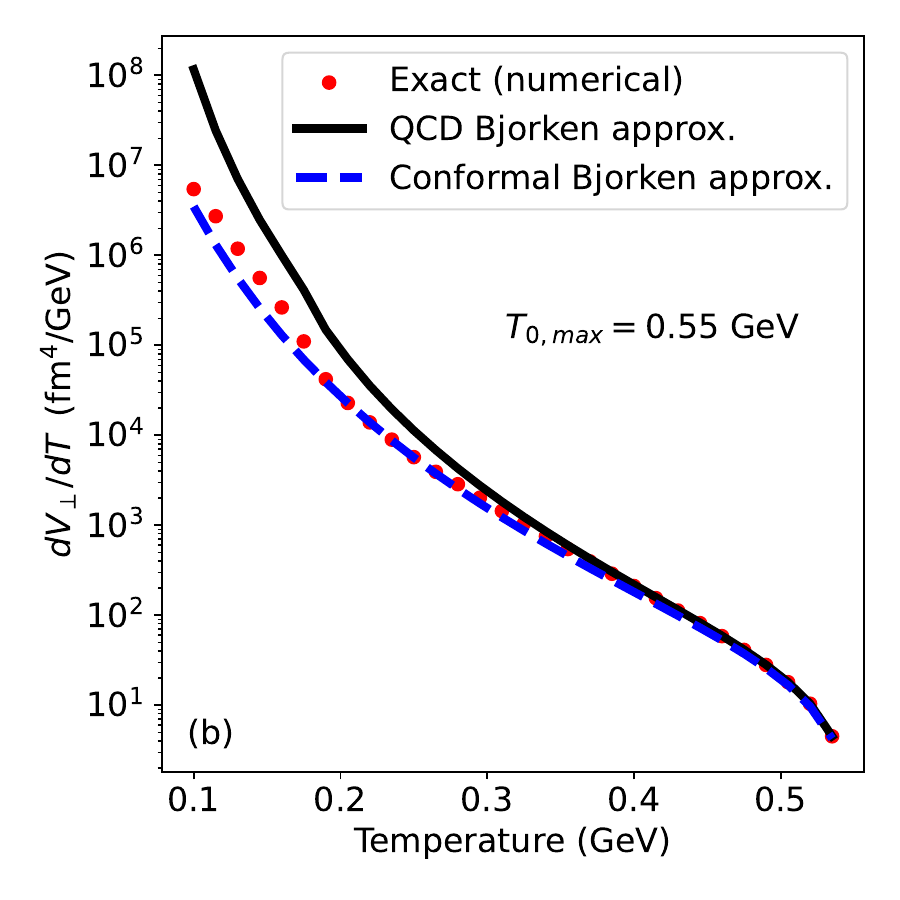}
	\includegraphics[width=0.72\linewidth]{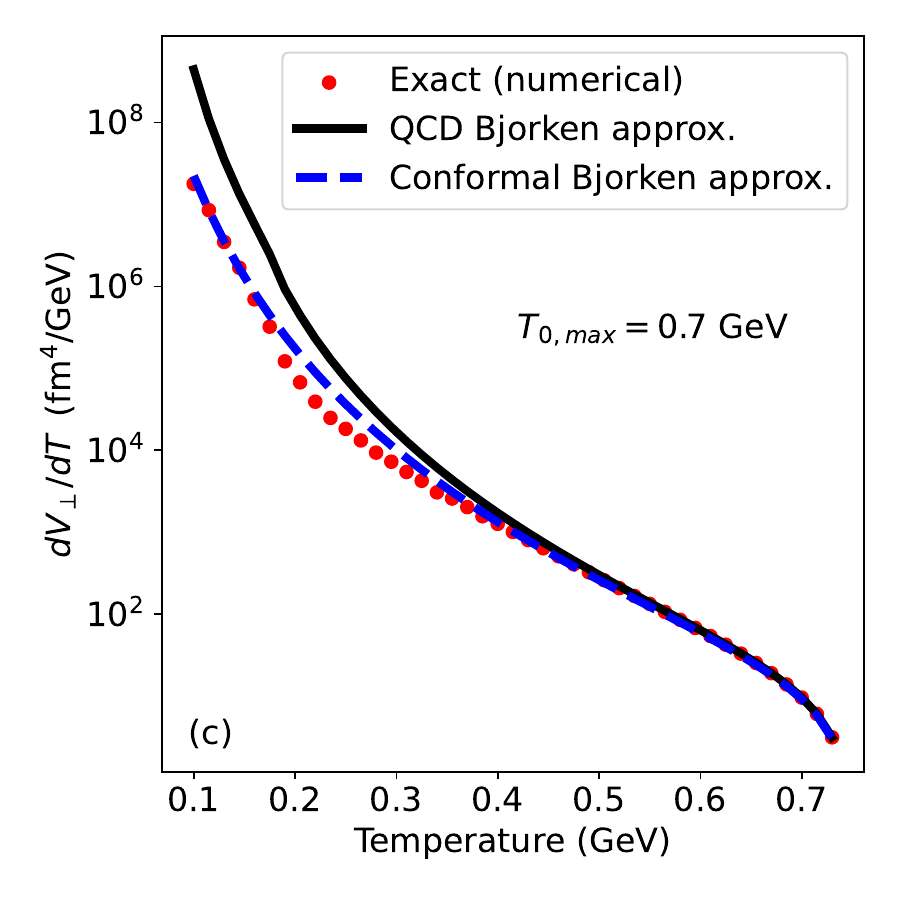}
	\caption{Transverse volume $d V_\perp/d T$ evaluated numerically (points) for ideal transversely-cylindrically-symmetric boost-invariant hydrodynamics with a Gaussian radial initial temperature profile set at time $\tau_0=0.4$~fm with a width of $\sigma_0=5$~fm and the QCD equation of state~\cite{Borsanyi:2013bia,Bernhard:2018hnz,eos_code},
	compared to Eq.~\ref{eq:dVdT_var_cs2} with the QCD equation of state (solid line) or with a conformal equation of state (dashed line), for three different maximum initial temperatures: (a) $\Tmax=0.3$~GeV; (b) $\Tmax=0.55$~GeV; (c) $\Tmax=0.7$~GeV. 
	Note that Eq.~\ref{eq:dVdT_var_cs2} with a constant speed of sound is equivalent to Eq.~\ref{eq:dVdT_gaussian_cs2}.}
	\label{fig:dVdT_num_vs_analytical}
\end{figure}

For purely coincidental reasons, however, the approximations are broken in a way that partly cancels their inaccuracy. At late times, transverse expansion leads to a \emph{faster} cooling of the plasma compared to a purely one-dimensional longitudinal expansion. On the other hand, at late times, the temperature of the plasma is lower, and its speed of sound is smaller too, reducing from $c_s^2 \approx 1/3$ to $c_s^2 \approx 1/7$. This reduction of the speed of sound at late times leads to a \emph{slower} cool down of the plasma compared to a plasma expanding with a constant $c_s^2=1/3$ rate.

This is relevant, because approximate analytical formulas can be derived for Bjorken expansion with a non-constant speed of sound, and exact solutions that account for transverse expansion are known as well ~\cite{Gubser:2010ze,Gubser:2010ui}; however, accounting \emph{simultaneously} for a realistic equation of state and transverse expansion has generally only been possible numerically.

This impacts the calculation of the transverse volume, Eq.~\ref{eq:dVdT_gaussian_cs2}. 
Again assuming a Gaussian initial temperature profile (Eq.~\ref{eq:T0_profile_gaussian}) and a longitudinal expansion, as assumed for Eq.~\ref{eq:dVdT_gaussian_cs2}, but generalizing for the scenario of a temperature-dependent speed of sound, 
a new approximate expression can be derived for the transverse volume:
\begin{align}
	\frac{d V_\perp}{ d T} \approx \frac{\pi  \sigma_0^2 \tau_0^2 \left( \left(\frac{\Tmax}{T}\right)^{2 c_s^{-2}\left(\sqrt{\Tmax T }\right)}-1\right)}{T} \frac{c_s^2\left( \sqrt{\Tmax T } \right)}{c_s^2\left(T\right)}
	\label{eq:dVdT_var_cs2}
\end{align}

If there were no accidental cancellation of errors between the speed of sound and the transverse expansion, one would expect Eq.~\ref{eq:dVdT_var_cs2} to be superior to Eq.~\ref{eq:dVdT_gaussian_cs2} when comparing with numerical simulations of  ideal transversely-cylindrically-symmetric boost-invariant hydrodynamics with the QCD equation of state.
In reality, simulations show that Eq.~\ref{eq:dVdT_var_cs2} is only an improvement over Eq.~\ref{eq:dVdT_gaussian_cs2} for systems where the evolution is dominated by Bjorken expansion for most of their lifetime, and the initial temperature of the system is sufficiently low that the QCD equation of state deviates significantly from $c_s^2=1/3$. This can be seen clearly in Figure~\ref{fig:dVdT_num_vs_analytical}: the exact transverse volume, evaluated numerically using a QCD equation of state, is compared to Eq.~\ref{eq:dVdT_var_cs2} with a QCD or conformal equation of state, for three different maximum initial temperature $\Tmax$:
\begin{itemize}
	\item  For the lowest initial temperature, $\Tmax=0.3$~GeV, the temperature is sufficiently low that the speed of sound of QCD deviates from $c_s^2=1/3$ even in the region where the plasma is effectively undergoing a one-dimensional Bjorken expansion. In that case, Eq.~\ref{eq:dVdT_var_cs2} with the QCD equation of state describes very well the exact calculations, only failing at lower temperatures where transverse expansion becomes more important. Using a conformal equation of state in Eq.~\ref{eq:dVdT_var_cs2} is clearly worse.
	\item  For a higher initial temperature, $\Tmax=0.55$~GeV (same as used in Figures~\ref{fig:flow_cs2_dep_and_approx}-\ref{fig:spectrum_with_without_flow}), the speed of sound at high temperatures is close to $c_s^2=1/3$, and using Eq.~\ref{eq:dVdT_var_cs2} with either a conformal equation of state or the QCD one is almost identical at high temperature. At low temperature, the accidental cancellation discussed above leads the conformal approximation to be significantly better than the QCD one.
	\item  For an even higher initial temperature, $\Tmax=0.7$~GeV, the same happens: using $c_s^2=1/3$ in Eq.~\ref{eq:dVdT_var_cs2} actually leads to a better description of the transverse volume.
\end{itemize}

Because the cancellation between the change in speed of sound and the transverse expansion is purely accidental, there will be different degrees of accuracy of Eq.~\ref{eq:dVdT_var_cs2} with $c_s^2=1/3$ or the QCD $c_s^2(T)$. However, values of $\Tmax$ around and above $0.55$~GeV are likely for central collisions at the Relativistic Heavy Ion Collider (top collision energy) and the Large Hadron Collider. Given the additional complexity of using $c_s^2(T)$, a good case can be made that it is only a clear benefit for lower energy collisions.

\section{Effect of Doppler shift on inverse slope $\Teff$ in other calculations}

\label{sec:Teff_prc}

 We investigated the effect of the transverse Doppler shift on the thermal photon spectra in Section~\ref{sec:effect_of_transverse_flow}. We found that the global effect of the Doppler shift varies depending on the range of energy where the inverse slope is extracted. We found that the effect of Doppler shift on the inverse slope is counter-intuitive: for $E \gtrsim 2$~GeV, the transverse Doppler shift reduces the inverse slope, contrary to the simple picture of the boosted rate
 \begin{multline}
 	k \frac{d^3 \Gamma_{\gamma}(E (\sqrt{1+u_\perp^2}-u_\perp),T)}{d^3 k}	\propto \exp\left(-\frac{E}{T \sqrt{\frac{1+v_\perp}{1-v_\perp}}}\right)
 \end{multline}
 where the transverse velocity increases the inverse slope. As we discussed, this is not a contradiction, since the measured photon spectrum is a highly non-trivial convolution of the above formula in space and time.

 \begin{table}[t]
 	\centering
 	
 	\begin{tabular}{|l|c|c|c|}
 		\hline
 		& \multicolumn{3}{|c|}{\textrm{Calculations from Ref.~\cite{Paquet:2015lta}}} \\
 		\hline
 		\textrm{$E$ range} & \textrm{w/ Doppler} & \textrm{w/o Doppler} & \textrm{rel. diff.} \\
 		 		\textrm{[GeV]} & \textrm{[GeV]} & \textrm{[GeV]} & \\
 		\hline
 		\textrm{full [0.2, 3.0]} & $0.295\,\pm\,0.012$ & $0.273\,\pm\,0.013$ & $+8\%$ \\
 		\hline
 		\textrm{0.2–1.0} & $0.189\,\pm\,0.017$ & $0.173\,\pm\,0.013$ & $+9\%$ \\
 		\hline
 		\textrm{1.0–2.0} & $0.315\,\pm\,0.005$ & $0.286\,\pm\,0.009$ & $+10\%$ \\
 		\hline
 		\textrm{2.0–3.0} & $0.369\,\pm\,0.003$ & $0.374\,\pm\,0.003$ & $-1\%$ \\
 		\hline
 	\end{tabular}
 	\caption{Extracted inverse slope $\Teff$ from $\frac{1}{2\pi E} \left.\frac{d N_{\rm thermal}}{d E dy_M}\right|_{y_M=0} \propto \exp(-E/\Teff)$ over different ranges of $E$, comparing Doppler-shifted vs non-shifted spectra. The right-most column is the relative difference between $\Teff$ with and without Doppler shift. The photon spectrum includes only thermal photons. See the text for details.}
 	\label{tab:Teff_ranges_150MeV_prc}
 \end{table}
 
 \begin{table}[t]
 	\centering
 	\begin{tabular}{|l|c|c|c|}
 		\hline
 		& \multicolumn{3}{|c|}{\textrm{$\Tmax$=0.55 GeV}} \\
 		\hline
 		\textrm{$E$ range} & \textrm{$\Teff$ w/ Doppler} & \textrm{$\Teff$ w/o Doppler} & \textrm{rel. diff.} \\
 		\textrm{[GeV]} & \textrm{[GeV]} & \textrm{[GeV]} & \\
 		\hline
 		\textrm{full [0.2, 3.0]} & $0.294\,\pm\,0.006$ & $0.278\,\pm\,0.008$ & $+6\%$ \\
 		\hline
 		\textrm{0.2–1.0} & $0.204\,\pm\,0.011$ & $0.184\,\pm\,0.009$ & $+11\%$ \\
 		\hline
 		\textrm{1.0–2.0} & $0.311\,\pm\,0.002$ & $0.292\,\pm\,0.005$ & $+6\%$ \\
 		\hline
 		\textrm{2.0–3.0} & $0.350\,\pm\,0.002$ & $0.370\,\pm\,0.002$ & $-5\%$ \\
 		\hline
 	\end{tabular}
 	\caption{Same quantities as in Table~\ref{tab:Teff_ranges_150MeV_prc}, but evaluated with the simpler inviscid hydrodynamics setting as used in Table~\ref{tab:Teff_ranges}, with $T_f$=150 MeV to match the calculations from Ref.~\cite{Paquet:2015lta}.}
 	\label{tab:Teff_ranges_inviscid_150MeV}
 \end{table}
 
 The results from Section~\ref{sec:effect_of_transverse_flow} used  a simpler model of heavy-ion collisions with ideal hydrodynamics and Gaussian initial conditions. In this Appendix, we confirm the generality of those results using a more sophisticated photon calculation that includes event-by-event initial conditions, viscous hydrodynamics with both shear and bulk viscosity and photon emission rates that interpolate between QGP and hadronic sources and include viscous corrections~\cite{Paquet:2015lta}. The photon spectra are for central Pb-Pb collisions at $\sqrts=2760$~GeV. Those same results had been used in Ref.~\cite{Paquet:2016ime} to investigate other aspects of the effect of the Doppler shift on the photon spectrum.
 
 The range of photon energy $E$ for which thermal photons were calculated in these older calculations is $0.2$ to $3$~GeV, which is smaller than the range investigated in  Section~\ref{sec:effect_of_transverse_flow}. Consequently, in the present section, smaller $E$ bins are used to compute the inverse slope. The maximum temperature in these simulations is close to $0.55$~GeV, corresponding approximately to the intermediate case ($\Tmax=0.55$~GeV) studied in Section~\ref{sec:effect_of_transverse_flow}.

The results are shown in Table~\ref{tab:Teff_ranges_150MeV_prc}. We see that the Doppler shift increases the inverse slope by approximately $10$\% for the low-$E$ photon spectrum, below $E=2$~GeV. For the photon bin $E \in [2,3]$~GeV, the transverse Doppler shift has no significant effect. This is in line with the results found in Section~\ref{sec:effect_of_transverse_flow}, where $E \approx 2$~GeV was the transition point between Doppler shift increasing or decreasing the inverse slope.

We can compare the inverse slopes extracted from the calculations from Ref.~\cite{Paquet:2015lta} with the simpler model (with $\Tmax=0.55$~GeV) from the present paper. We recomputed the inverse slope from Section~\ref{sec:effect_of_transverse_flow} using the same $E$ bins as Table~\ref{tab:Teff_ranges_150MeV_prc}. The results are presented in Table~\ref{tab:Teff_ranges_inviscid_150MeV}. The inverse slopes from the simple model are largely consistent with the more sophisticated calculations, and the general effect of the Doppler shift is also consistent. This provides evidence that the simpler model used in Section~\ref{sec:effect_of_transverse_flow} captures much of the physics relevant to understanding the inverse slope.

\bibliographystyle{apsrev4-2}
\bibliography{biblio}

\end{document}